\documentclass[a4paper,11pt]{article}
\usepackage{jcappub} 
\usepackage{lineno}
\usepackage{gensymb}
\usepackage{amsmath}
\title{\boldmath Two-component $\gamma$-Ray Structure from the CR Sources Within Dense Clouds}

\author[a,b]{Lin Nie}
\author[b,c,d]{Yi-Qing Guo}
\author[a]{Si-Ming Liu}
\affiliation[a]{School of Physical Science and Technology, Southwest Jiaotong University,\\ Chengdu, 610031, China\\}
\affiliation[b]{Key Laboratory of Particle Astrophysics,\\
Institute of High Energy Physics, Chinese Academy of Sciences,\\
Beijing, 100049, China}
\affiliation[c]{University of Chinese Academy of Sciences,\\ 
Beijing 100049, China }
\affiliation[d]{TIANFU Cosmic Ray Research Center,\\
Chengdu 610000, China}
\emailAdd{guoyq@ihep.ac.cn}
\emailAdd{liusm@swjtu.edu.cn}
\abstract{Recent observations have revealed that several cosmic ray (CR) sources themselves exhibit pronounced double power-law features in their radiation spectra. Combined with the phenomenon of two-component structure in the observed CR energy spectrum supported by multi-messenger data, this raises a fundamental question: can the two-component structure of the cosmic ray energy spectrum and the double power-law feature of the gamma-ray radiation energy spectrum from supernova remnants be understood within a unified picture? In this study, we propose a two-component model 
that incorporates the re-acceleration of background ``sea" CR particles by astrophysical sources to systematically explain the formation of double power-law spectra within those sources. Our model successfully reproduces the gamma-ray observations of multiple CR sources.  The results support that double power-law structures may be a generic feature of Galactic CR sources within crushed clouds.
This work offers a new theoretical perspective on the origin and propagation of cosmic rays, and its predictions may be further tested with future observations of a larger sample of CR sources.
}


\begin{document}
\maketitle
\flushbottom

\section{Introduction}
\label{sec:intro}
Since the discovery of cosmic rays (CRs) in 1912, understanding their acceleration mechanisms, origins, and propagation processes has remained a central pursuit in cosmic ray physics. Among these aspects, the propagation of CRs is widely regarded as the key to unveiling their origins. Over the past decade, significant advances in observational techniques have revealed a wealth of new spectral features, including noticeable changes in spectral indices at specific energy ranges and distinct cutoff behaviors among different nuclear species \citep{2015PhRvL.114q1103A,2015PhRvL.115u1101A,2017PhRvL.119y1101A,2018PhRvL.120b1101A,2018PhRvL.121e1103A,2020PhRvL.124u1102A,2021PhRvL.126d1104A}. These phenomena markedly deviate from the predictions of traditional CR propagation models, prompting a re-examination of both CR acceleration and transport mechanisms.

It is well established by AMS-02 and other experiments that cosmic ray energy spectra deviate from a simple power law, exhibiting a spectral hardening above several hundred GeV that has been independently confirmed by multiple measurements \citep{2021PhRvL.126d1104A,2019SciA....5.3793A,2011Sci...332...69A,2009BRASP..73..564P}. Moreover, secondary CR species also exhibit spectral hardening in the same energy range, and intriguingly, they are harder than the spectra of their primary counterparts \citep{2018PhRvL.120b1101A}. Two primary hypotheses have been proposed to account for this behavior: (1) a change in the source injection spectrum at high energies, or (2) a transition in propagation characteristics, especially a shift in the rigidity dependence of the diffusion coefficient due to varying turbulence properties of the interstellar medium. Given that the ratios of secondary to primary CRs, such as B/C and B/O, also display hardening at the same energies \citep{2016PhRvL.117i1103A,2022SciBu..67.2162D}, the latter explanation involving modified propagation has gained broader support.

To interpret these observations, we previously proposed a spatially dependent cosmic ray propagation model that successfully reproduces several key spectral features \citep{2018PhRvD..97f3008G,2024PhRvD.109f3001Y,2024ApJ...974..276N,2023ApJ...956...75Q}. In this model, The halos of CR sources embedded within the Galactic disk confine local particles and impede the passage of CRs from more distant sources due to the smaller diffusion coefficient around them. As a result, CRs from remote sources can only reach the other regions via diffusion along the Galactic outer halo. The model predicts that low-energy CRs (below ~100 GeV) predominantly originate from a spatially stable background ``sea" of particles distributed across the Galaxy, while high-energy CRs are mainly contributed by local sources, with their properties varying spatially. Furthermore, the model anticipates that the diffuse gamma-ray emission throughout the Galaxy should commonly exhibit a double power-law structure.

Interestingly, although the gamma-ray spectra of Galactic CR sources show considerable diversity, recent observations have revealed that many sources within the crushed clouds, particularly supernova remnants (SNRs), exhibit pronounced double power-law features \citep{2025ApJ...982L..33C,2023Univ....9...98S,2021MNRAS.502..472A,2025arXiv250305261N}. This raises the possibility of an additional spectral component of different origin within these sources. A natural question thus arises: could this component be associated with the background “sea” of Galactic CRs? While background subtraction is routinely employed to isolate intrinsic source emission in gamma-ray data analysis, limitations in detector resolution often make complete removal of the background challenging. Moreover, recent studies suggest that CRs permeating the cloud can be re-accelerated and compressed at the SNR shock
\citep{2010ApJ...723L.122U,2015ApJ...800..103T,2015ApJ...806...71L,2016A&A...595A..58C}, thus providing an additional distribution. 

In this work, we propose a two-component model potentially applicable to a wide range of Galactic CR sources. By incorporating the reacceleration and compression of background ``sea" CRs by source-induced shocks, we develop a coherent physical picture that accounts for the widespread presence of double power-law features in CR energy spectra, diffuse gamma-ray emission, and radiation spectra of individual sources. The structure of the paper is as follows: Section \ref{sec:method} outlines the methodology, Section \ref{sec:result} presents and discusses the results, and Section \ref{sec:conclusion} concludes the study with a summary and outlook for future research.

\section{MODEL \& METHODOLOGY} \label{sec:method}

We propose that the high-energy radiation from cosmic ray sources consists primarily of two components: the first arises from freely injected thermal particles accelerated by the source, producing radiation via inverse Compton scattering (for electrons) or through proton-proton (pp) interactions; the second originates from background “sea” cosmic rays that are re-accelerated by local diffusive shocks or bow shocks, and subsequently interact with the ambient gas. The former dominates the high-energy emission, while the latter accounts for the low-energy emission. 

In order to calculate the contribution of re-acceleration or adiabatic compression of pre-existing local cosmic rays from various sources to their observed gamma-ray radiation, it is crucial to compute the spatial distribution of galactic cosmic rays.

\subsection{Calculation of Background ``Sea" CR protons}
After being accelerated and escaping from their sources, background cosmic ray (CR) particles propagate through the Galaxy by diffusing in the interstellar magnetic field. Their distribution can be affected by convection and reacceleration due to scattering with random magnetohydrodynamic waves. Additionally, interactions between CRs and interstellar gas, radiation fields, and magnetic fields result in spallation and energy losses. The process by which cosmic rays fill the entire Galaxy through diffusion after injection from their sources is highly complex.

The key point is the diffusion coefficient in this model. Here, it is assumed to be anti-correlated with the CR source distribution, which is a natural assumption since the (turbulent) magnetic field strength is expected to be correlated with the matter distribution. This means that in regions with more cosmic ray sources, there is a smaller diffusion coefficient. In fact, this assumption is supported by observational data \citep{2017Sci...358..911A,2025PhRvL.134q1005A}. For spatially dependent propagation (SDP) model adopted in this work, since the regions surrounding cosmic ray sources are treated as slow diffusion zones, the propagation of cosmic rays through the galaxy exhibits: on the one hand, the halos of cosmic ray sources embedded within the Galactic disk hinder the propagation of incoming cosmic-ray particles due to smaller diffusion coefficient; on the other hand, they also confine and capture local cosmic-ray particles. 
The propagation of cosmic rays is precisely described by the following equation \citep{2007ARNPS..57..285S}:
\begin{equation}
\label{Eq2.1}
\begin{aligned}
\frac{\partial \psi(\vec{r}, p, t)}{\partial t}= & q(\vec{r}, p, t)+\vec{\nabla} \cdot\left(D_{xx} \vec{\nabla} \psi-\vec{V} \psi\right) \\
& +\frac{\partial}{\partial p} p^2 D_{pp} \frac{\partial}{\partial p} \frac{1}{p^2} \psi-\frac{\partial}{\partial p}\left[\dot{p} \psi-\frac{p}{3}(\vec{\nabla} \cdot \vec{V}) \psi\right] \\
& -\frac{1}{\tau_f} \psi-\frac{1}{\tau_r} \psi,
\end{aligned}
\end{equation}

Here, $\psi(\vec{r}, p, t)$ represents the density of cosmic rays with momentum $\rm p$ at position $\vec{r}$, and $q(\vec{r}, p, t)$ denotes the source term. $\rm D_{xx}$ is the spatial diffusion coefficient, $\vec{V}$ is the convection velocity, $\rm D_{pp}$ is the diffusion coefficient in momentum space, $\rm \tau_{f}$ is the timescale for fragmentation, and $\rm \tau_{r}$ is the timescale for radioactive decay.


The source term $q(\vec{r}, p, t)$ is used to describe the spatial distribution of cosmic ray sources and their injection power spectrum. It characterizes the spatial distribution features of cosmic ray sources and their injection momentum spectrum. As a primary source of galactic cosmic rays, the spatial distribution of supernova remnants is widely employed to represent the distribution pattern of galactic cosmic ray sources. In cylindrical coordinates, this distribution can be expressed as \citep{1996A&AS..120C.437C}
\begin{equation}
f(r, z) \propto\left(r / r_{\odot}\right)^{1.69} \exp \left[-3.33\left(r-r_{\odot}\right) / r_{\odot}\right] \exp \left(-|z| / z_s\right)
\end{equation}
Where $\rm r_{\odot}$ = 8.5 kpc and $\rm z_s$ = 0.2 kpc. The coupling relationship of the $\rm D_{pp}$ with the spatial diffusion coefficient $\rm D_{xx}$ is given by:
\begin{equation}
D_{\mathrm{pp}} D_{\mathrm{xx}}=\frac{4 p^2 v_A^2}{3 \delta\left(4-\delta^2\right)(4-\delta)w}
\end{equation}
where $\rm v_A$ is the Alfvén speed, and $\rm w$ is the ratio of magneto-hydrodynamic wave energy density to the magnetic field energy density, which can be fixed to 1.

Similar to previous studies \citep{2018PhRvD..97f3008G,2024PhRvD.109f3001Y,2024ApJ...974..276N,2025arXiv250305261N}, to appropriately account for the observed characteristics of cosmic rays, we consider a spatially dependent propagation mechanism for them. The cosmic-ray diffusion depends on the distribution of cosmic-ray sources $\rm f(r, z)$, and the
diffusion coefficient is described as
\begin{equation}
D_{x x}(r, z, \mathcal{R})=D_0 F(r, z) \beta^\eta\left(\frac{\mathcal{R}}{\mathcal{R}_0}\right)^{\delta_0 F(r, z)},
\label{eq4}
\end{equation}
where $\rm \delta_0 F(r, z)$ describes the turbulent characteristics of the local medium environment and $\rm D_0 F(r, z)$ represents the normalization factor of the diffusion coefficient at the reference rigidity.
\begin{equation}
    F(r, z)= \begin{cases}g(r, z)+[1-g(r, z)]\left(\frac{z}{\xi z_0}\right)^n, & |z| \leq \xi z_0 \\ 1, & |z|>\xi z_0\end{cases},
\end{equation}
here, $\rm \xi z_0$ denotes the half-thickness of the Galactic halo, and $\rm g(r, z)=N_m /[1+f(r, z)]$, where $\rm N_m$ is the normalization factor. The parameter $\rm n$ is used to describe the smoothness between the Galactic inner and outer halo.

We assume a power-law spectrum in rigidity for the injected cosmic rays (CRs). Employing the GALPROP numerical package \citep{1998ApJ...509..212S}, we solve the CR propagation equation to derive their Galactic distribution.

\begin{figure}[t]
\centering
\includegraphics[width=0.95\linewidth]{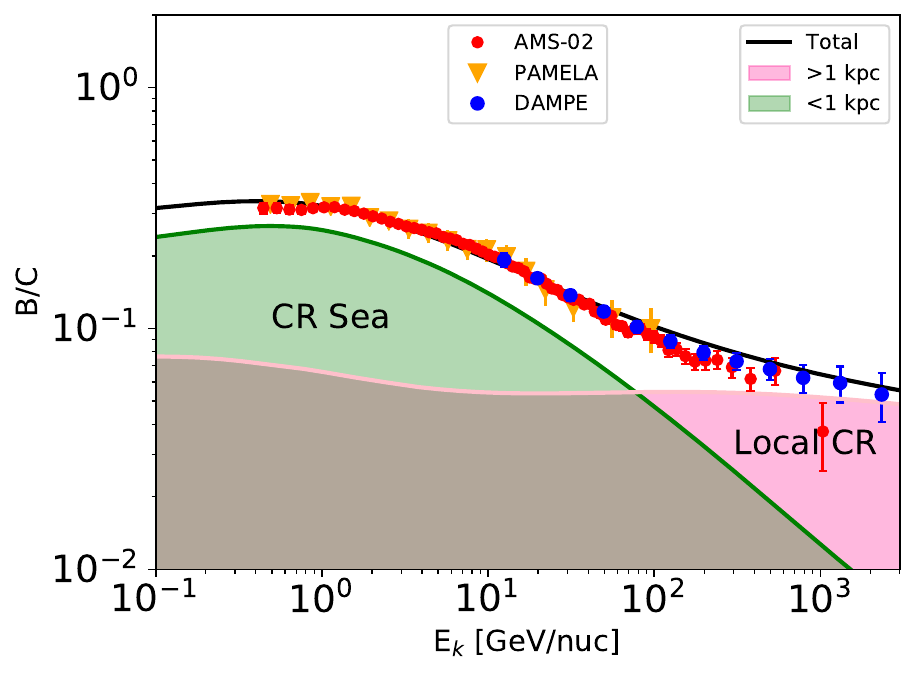}
\caption{A comparison of the B/C ratio calculated using the CR SDP model with observational data from AMS-02 \citep{2017PhRvL.119y1101A}, PAMELA \citep{2014ApJ...791...93A}, and DAMPE \citep{2022SciBu..67.2162D}. The pink and green shaded areas represent the fluxes from sources within ($\rm r < 1 ~kpc$) and beyond ($\rm r > 1 ~ kpc$) 1 kpc, respectively. In this study, the solar modulation potential is consistently assumed to be 550 MeV.}
\label{fig1}
\end{figure}
\begin{figure}[t]
    \centering
    \includegraphics[width=0.95\linewidth]{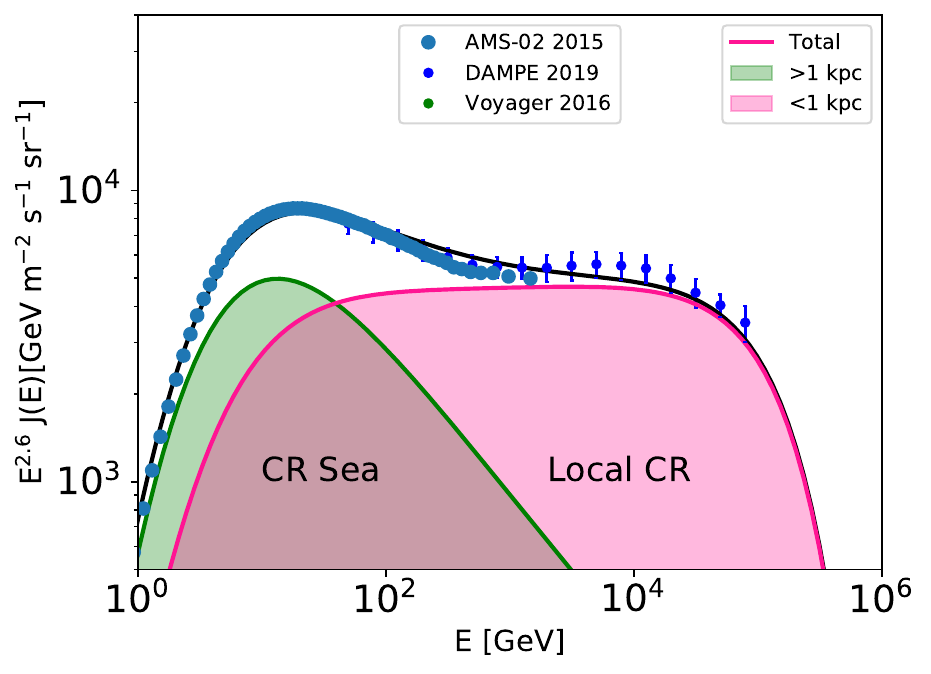}
    \caption{The CR proton spectrum is calculated using the CR SDP model and compared with observational data from AMS-02 \citep{2015PhRvL.114q1103A} and DAMPE \citep{2019SciA....5.3793A}. The pink and green shaded areas represent the fluxes from sources within ($\rm r < 1~kpc$) and beyond ($\rm r > 1 ~kpc$) 1 kpc, respectively.}
    \label{fig2}
    \end{figure}

\begin{table}[htbp]
\centering
\fontsize{12}{20}\selectfont
\caption{Parameters of the SDP model.}
\label{tab1}
\setlength{\tabcolsep}{14pt}
\begin{tabular}{lcccccc}    
\hline
$D_0{ }\left[\mathrm{cm}^{2} \mathrm{~s}^{-1}\right]$ & $\delta_0$ & $N_m$ & $\xi$ & $\mathrm{n}$ & $v_A\left[\mathrm{~km} \mathrm{~s}^{-1}\right]$ & $z_0[\mathrm{kpc}]$ \\
\hline $4.8 \times 10^{28}$ & 0.63 & 0.24 & 0.1 & 4.0 & 8 & 4.5 \\
\hline 
\end{tabular}
\end{table}

\begin{figure*}[htbp]
\centering
\includegraphics[width=0.48\linewidth]{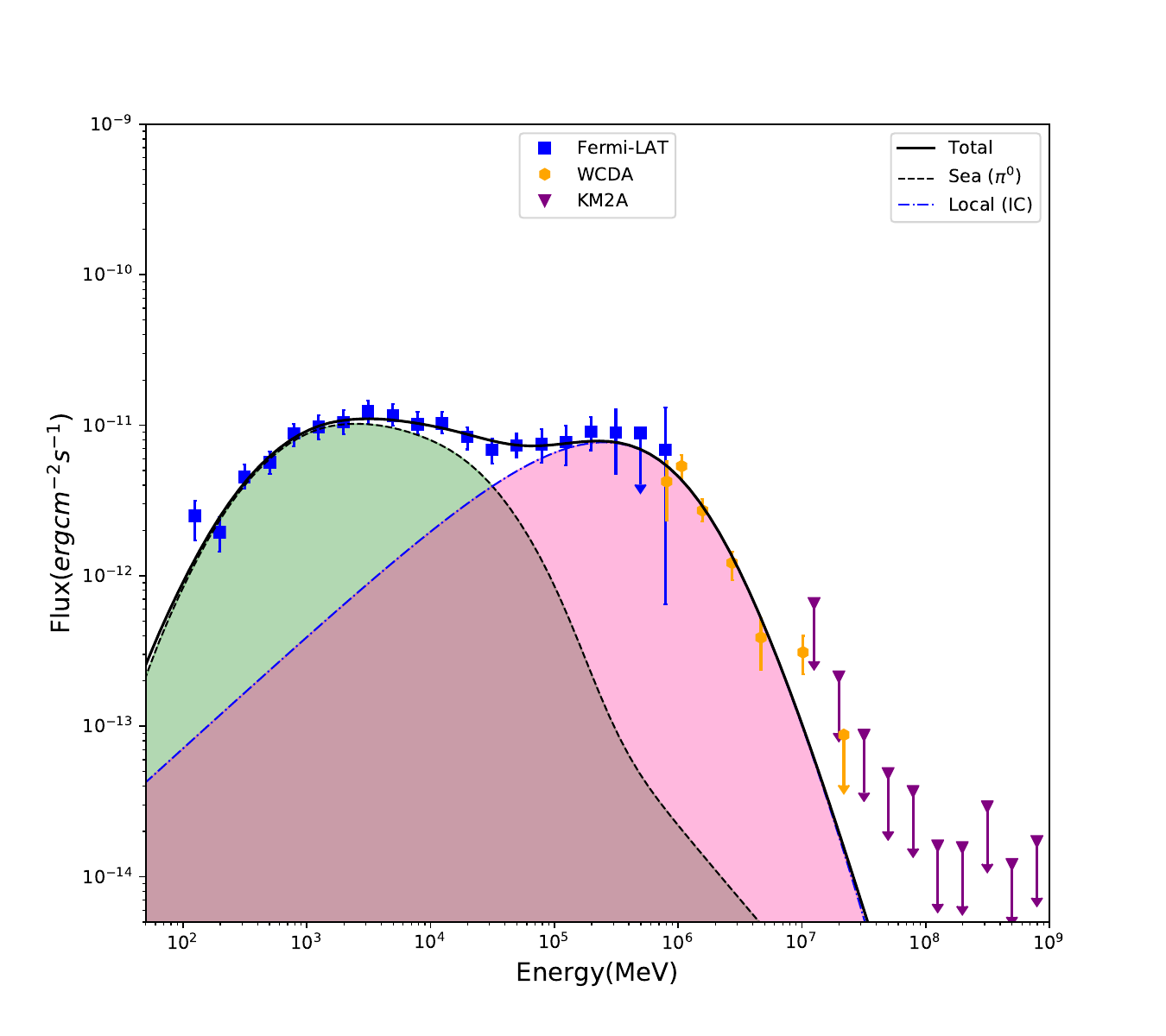}
\includegraphics[width=0.48\linewidth]{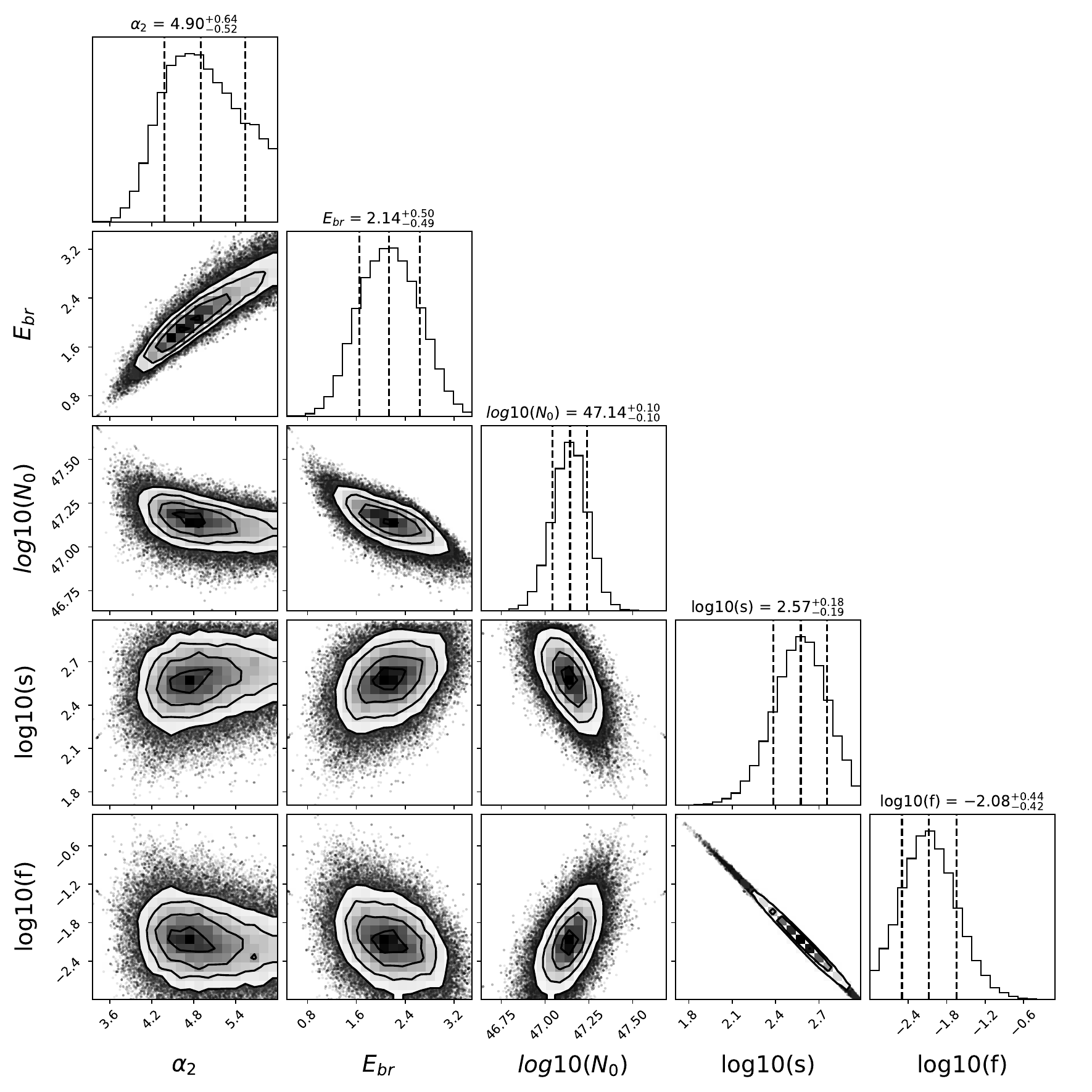}
\caption{The gamma ray spectra predicted by model compared with the observational data for SNRs CasA (left). The observational data are taken from the LHAASO and Fermi-LAT \citep{2025ApJ...982L..33C}. The right panel show that the fitting 1-dimensional probability distributions (diagonal) and 2-dimensional credible regions of the model parameters.} 
\label{fig3}
\end{figure*}

\subsection{Reacceleration and Adiabatic Compression of Background ``Sea" CRs}
Given their pre-existing non-thermal state, CRs within the cloud face no injection barrier and are reaccelerated at the SNR shock regardless of their energy. Here, we compute the reaccelerated spectra downstream of the shock by adopting the formalism of \cite{2004APh....21...45B}, which was originally developed for the more complex case of non-linear shock reacceleration.
The distribution of cosmic ray particles upstream and downstream of the shock is described by the following equation: 
\begin{equation}
u \frac{\partial f}{\partial x}-\frac{\partial}{\partial x} \kappa \frac{\partial f}{\partial x}=\frac{1}{3} \frac{d u}{d x} p \frac{\partial f}{\partial p} .
\label{eq2}
\end{equation}
where $\rm f(x, p)$ is the CR distribution function, u is shock velocity and $\rm k(p)$ is the energy dependent diffusion coefficient. The $\rm x$ axis is parallel to the shock normal and the shock is located at $\rm x = 0$. The boundary condition at upstream infinity (located at $\rm x = -\infty$) is imposed by requiring that the distribution function $\rm f(-\infty,p)$ be equal to the Galactic background ``sea" CR distribution $\rm f_{GCR}(p)$. The solution of the transport equation at the shock location, $f_{acc}(p) = f(x = 0, p)$, is easily found in the form \citep{1978ApJ...221L..29B}:
\begin{equation}
f_{acc}(p)=\alpha\left(\frac{p}{p_m}\right)^{-\alpha} \int_{p_m}^p \frac{d p^{\prime}}{p^{\prime}}\left(\frac{p^{\prime}}{p_m}\right)^\alpha f_{GCR}\left(p^{\prime}\right)
\end{equation}
where $\rm \alpha = \frac{3v_{sh}}{v_{sh}-u_d}$ and $u_d$ is the gas velocity downstream of the shock. The momentum $p_m$ represents a minimum momentum in the spectrum of Galactic CRs. The significance of this reacceleration is twofold. Notably, it elevates the momentum per particle, with the maximum momentum determined by the balance between the acceleration timescale and the system age. Furthermore, it leads to spectral hardening, but only if the Galactic CR spectrum is steeper than the $\rm p^{-\alpha}$. Otherwise, the spectral index remains unchanged, and only the normalization differs.

\begin{figure}[h]
\centering
\includegraphics[width=0.98\linewidth]{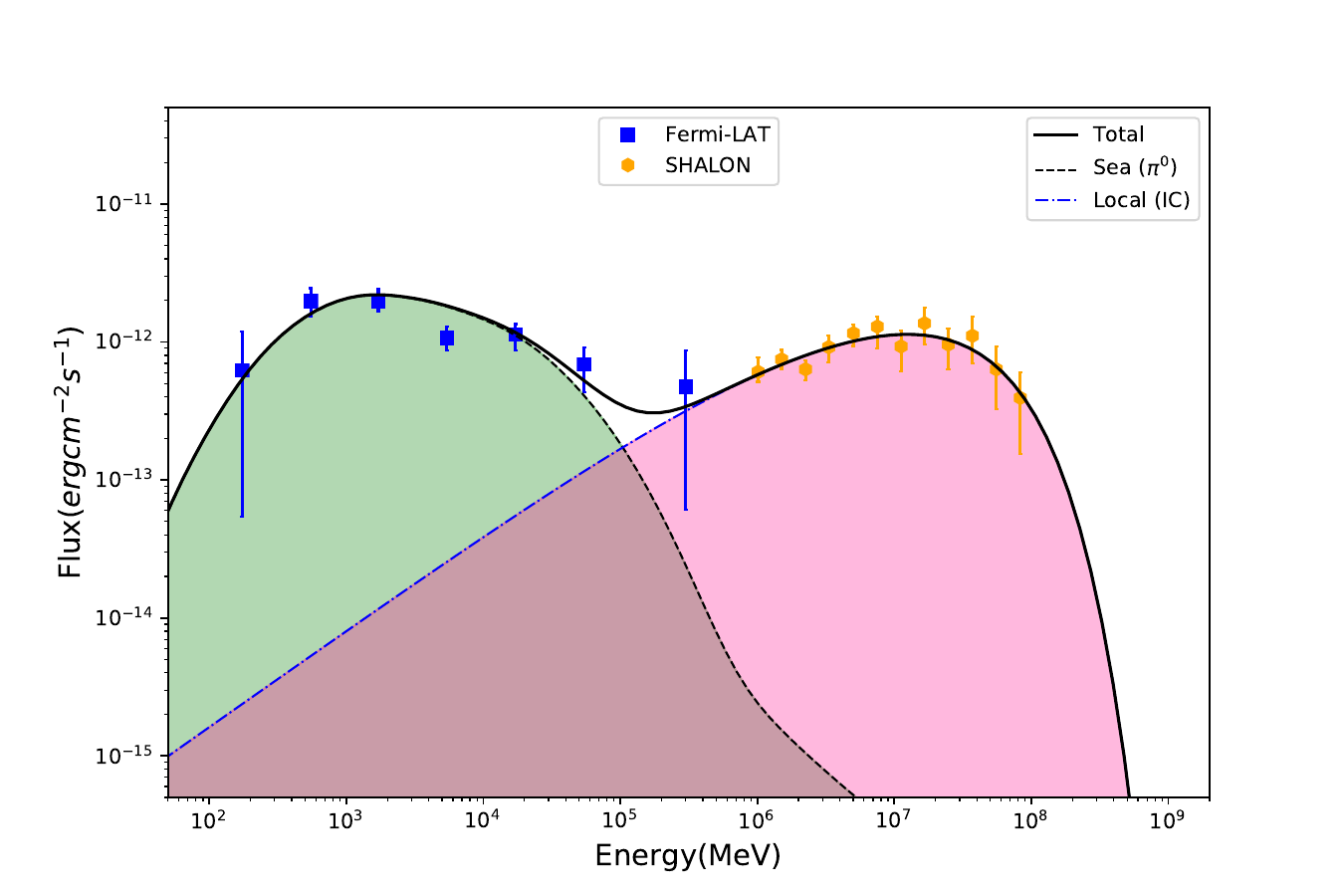}
\caption{The same as left panel of figure \ref{fig3} but for the Tycho. The observational data are taken from the Fermi-LAT \citep{2022ApJS..260...53A} and SHALON \citep{2018AdSpR..62.2845S}.}
\label{fig4}
\end{figure}

\begin{figure*}[htbp]
    \centering
    \includegraphics[width=0.48\linewidth]{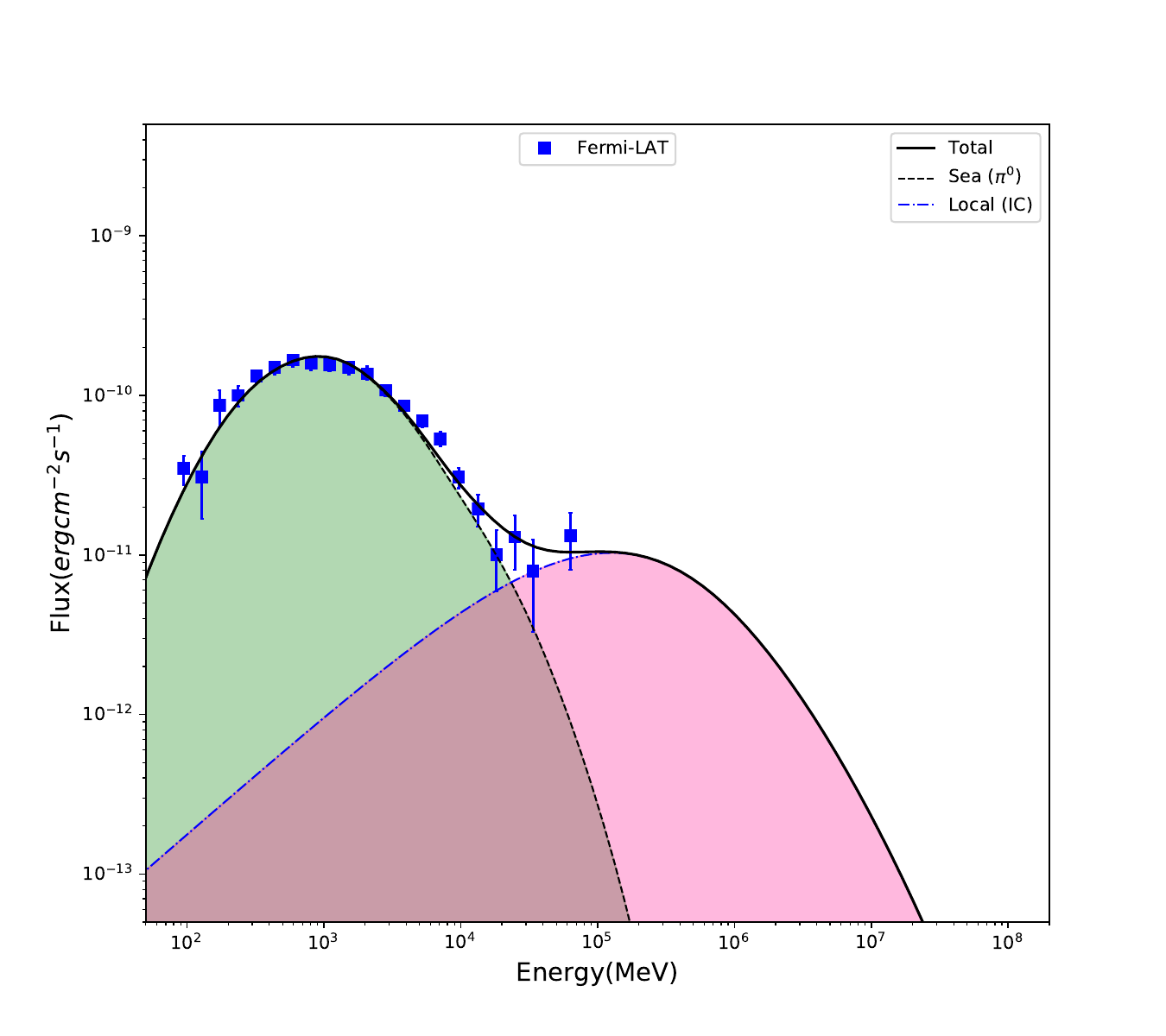}
    \includegraphics[width=0.48\linewidth]{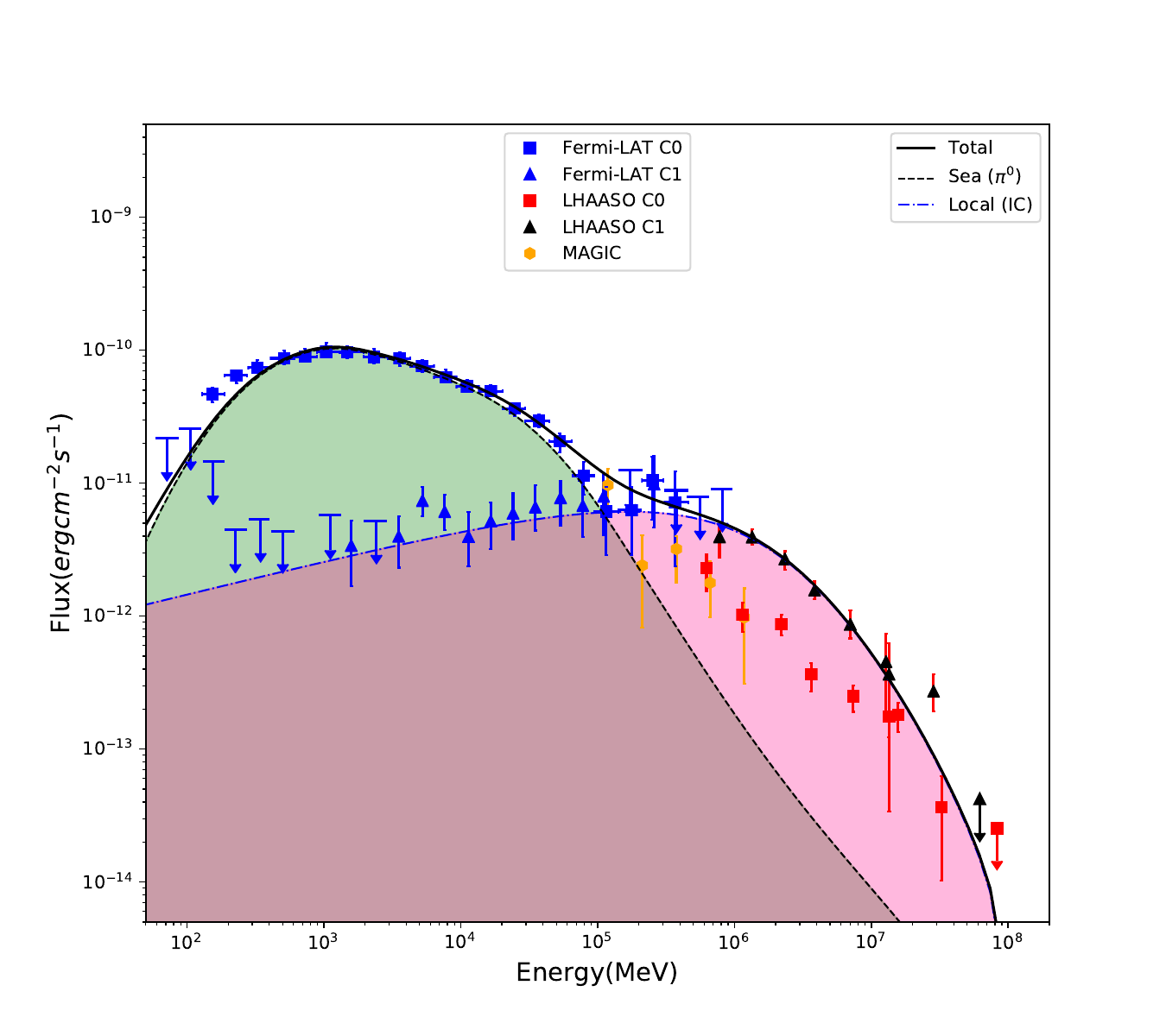}
    \caption{The same as left panel of figure \ref{fig3} but for the W44 (left panel) and IC443 (right panel). The observational data are taken from the Fermi-LAT \citep{2013Sci...339..807A}, MAGIC\citep{2007ApJ...664L..87A} and LHAASO \citep{2025arXiv251026112C}. Here, C0 and C1 refer to the regions around a pointlike source, with angular extensions of approximately $0.27\degree$ and $0.67\degree$, respectively \citep{2025arXiv251026112C}.}
    \label{fig5}
    \end{figure*}

\begin{table}[htbp]
\centering
\fontsize{12}{16}\selectfont
\setlength{\tabcolsep}{4pt}
\label{tab2}
\begin{tabular}{lcccccccccc}    
\hline 
Name & $\substack{\rm d \\ \rm (kpc)}$ & $\substack{\rm Age \\ \rm (year)}$ & $\substack{\rm n_{0}\\(\mathrm{cm}^{-3})}$ & $\substack{\rm p_{br} \\ \rm (GeV/c)}$ & $\rm log_{10}(s)$ & $\rm log_{10}(f)$ & $\substack{\rm log10(N_0) \\ (\rm TeV^{-1})}$ & $\substack{\rm E_{br} \\ \rm (TeV)}$ & $\alpha_2 $ & ref \\
\hline
CasA & 3.4 & 350 & 4.4 & 40 & 2.56 & -2.09 & 47.2 & 21.4 & 4.90 & \citep{2025ApJ...982L..33C,2010ApJ...720...20A} \\
Tycho & 2.5-3.0 & 440 & 0.6 & 40 &1.9 & -0.7 & 45.22 & 71.4 & 3.84 & \citep{2012ApJ...744L...2G,2012ApJ...761..133Y} \\
W44 & 2.9 & 10000 & 200.0 & 20 & 0.97 & -1.08 & 47.41 & - & - & \citep{2005ApJ...618..297R,2010ApJ...723L.122U} \\
IC443 & 1.5 & 30000 & 15 & 50 & 0.94 & 0.06 & 46.3 & 120 & 3.8 & \citep{2010ApJ...712..459A,2013Sci...339..807A,1980ApJ...242.1023F} \\
\hline 
\end{tabular}
\caption{Parameters of the sources and the best fitting value of the relevant model parameters. With the exception of IC443, the free parameter $\alpha_1$ for all radiation sources is set to 1.5; whereas for IC443, the value of $\alpha_1$ is constrained to 2.5 based on observational data. Additionally, the cutoff energy $\rm E_{cut}$ is fixed at 100 TeV for all sources.}
\end{table}

\begin{figure}[h]
\centering
\includegraphics[width=0.98\linewidth]{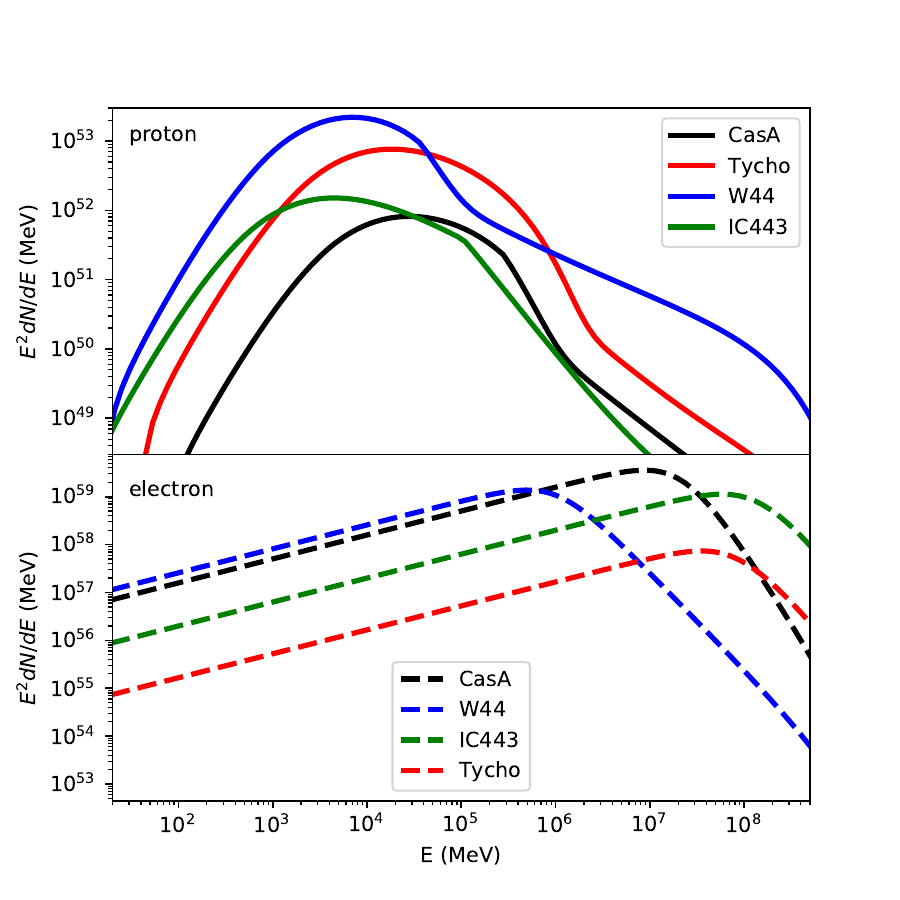}
\caption{The top panel shows the volume collected spectrum of re-accelerated CR protons as described by Equation (\ref{eq8}), while the bottom panel corresponds to the electron spectrum given by Equation (\ref{eq12}). The corresponding parameters, which are constrained by the observed $\gamma$-ray data, are listed in Table \ref{tab2}.}
\label{fig6}
\end{figure}

For a slow shock running into a dense medium, the weak ionization in the shock precursor enhances ion-neutral damping of Alfvén waves. This allows high-energy cosmic-ray particles to escape the acceleration site, leading to a steepening of the particle spectrum by a factor of $\rm p_{br}/p$ above the break momentum $\rm p_{br}$.

The high-energy particles accelerated at the shock are further heated by adiabatic compression as the gas density increases, until magnetic pressure provides support. The interaction between an SNR shock and a dense cloud leads to the formation of a radiative shell behind the shock front. The subsequent adiabatic compression of the gas within this shell raises its density to very high levels, which can significantly boost gamma-ray emission from pion decay. This compression also re-energizes pre-existing Galactic cosmic rays trapped in the shell and increases their overall density, thereby enhancing the total cosmic ray spectrum.  Due to this further compression, the number density of accelerated and compressed CRs is \citep{1982ApJ...260..625B}:
\begin{equation}
f_{ad}(p)=s^{2 / 3}f_{acc}\left(s^{-1 / 3} p\right)
\label{eq8}
\end{equation}
where $\rm s$ is the compression factor is equal to
\begin{equation}
s \equiv\left(\frac{n_m}{n_d}\right)=\left(\frac{n_m}{r_{s h} n_0}\right)
\end{equation}
where $\rm n_d$ is density immediately downstream of the shock, $\rm n_m/n_d$
is compression ratio due to radiative cooling and $\rm r_{sh} = n_d/n_0$ is the shock compression ratio due to the shock that we have already seen.

In order to obtain the final particle spectrum in compressed clump,  We solve a usual kinetic equation to obtain re-accelerated proton spectrum,
\begin{equation}
\frac{\partial N_p(p, t)}{\partial t}=\frac{\partial}{\partial p}\left[b(p) N_p(p, t)\right]+Q_p(p),
\end{equation}
This equation describes the temporal evolution of the re-accelerated CR proton spectrum in the compressed region. Here, $b(p)$ represents proton energy losses, and $Q(p)$ denotes the proton injection rate, expressed as
\begin{equation}
Q(p)=\frac{n_0}{n_m t_c} f_{\mathrm{ad}}(p), 
\end{equation}
Here, $t_c$ is the interaction time between the cloud and the source.

\subsection{Component of the Thermal Injected Particles}
Initially, the fresh acceleration of particles at the shock was proposed as the primary source of $\gamma$ -ray emission; it continues to be regarded as a significant contributor. Given that $\gamma$-ray emission from the SNR in low-density environments is inverse-Compton dominated \citep{2012ApJ...761..133Y}, we model this freshly accelerated component as inherently leptonic, characterized by a broken power-law distribution with an exponential cutoff,
\begin{equation}
N(E)=N_0\left(\frac{E}{1 \mathrm{TeV}}\right)^{-\alpha_1} \left[1+\left(\frac{E}{E_{\mathrm{br}}}\right)^2\right]^{\frac{\alpha_1-\alpha_2}{2}}\exp \left(\frac{-E}{E_{\mathrm{cut}}}\right)
\label{eq12}
\end{equation}
where $\rm \alpha_1$ and $\rm \alpha_2$ are spectral indices below and after $E_{br}$. The $\rm N_0$ is the normalization factor, The cut-off energy $\rm E_{cut}$ is the maximum energy for which particles can be accelerated.

\section{Results and Discussion} \label{sec:result}
In this section, we present our computational results. To calculate the composition of gamma rays produced by background ``sea" CRs which are re-accelerated by the process of diffusive shock acceleration and experience further heating due to adiabatic compression in crushed cloud regions, we utilize the data of cosmic ray B/C and protons observed near Earth to constrain the relevant parameters of the cosmic ray propagation model, thus determining the proton distribution of cosmic rays at different spatial locations. Figure \ref{fig1} and Figure \ref{fig2} respectively show the fitting results of the B/C and proton observational data. The model parameters are presented in Table \ref{tab1}.
As shown in Figure \ref{fig1} and \ref{fig2}, the high-energy cosmic rays originate from local cosmic ray sources in local regions, while the low-energy cosmic rays result from the unified contribution of accelerated and propagated cosmic rays from distant sources.

Given the above particle distribution and density, we employ the PYTHON package Naima \citep{2015ICRC...34..922Z} to calculate the $\gamma$-ray spectra.
When we calculate the non-thermal radiation from molecular clouds, the gas density in the region is considered constant, and high-energy particles are distributed throughout the volume, with their energy spectrum integrated over the volume $\rm V$, where $\rm V=4\pi R^3/3 $ is the SNR volume with a radius R. The volume of the preshock cloud is then $\rm f V$ , the $\rm f$ is the preshock filling factor of the compressed cloud.

We fit the gamma-ray spectra of different sources by adjusting several free parameters, including the compression factor $\rm s$, the preshock filling factor of the compressed cloud $\rm f$, and the injection spectrum parameters $\alpha_2$, $N_0$ and break energy $\rm E_{b}$.
Here, we use the MCMC sampling method \citep{2013PASP..125..306F} to derive the best-fitting values of the above parameters.
Figure \ref{fig3} shows calculated results of the young supernova remnants (SNRs) Cas A that include the distribution of parameters, and results of comparison between the theoretical spectrum with the best-fitting model parameters and observed data.
Our model successfully reproduces its observed spectra. It exhibits comparable double power-law features. As representative examples, it is located in high-density environments where the GeV-band emission is primarily produced by reaccelerated background CRs interacting with ambient gas, while the TeV emission is dominated by radiation from freshly accelerated electrons.

The density of the target gas around Cas A is taken as $\rm 3~ cm^3$, which has been derived by investigating the infrared emission \citep{2008ApJ...678..287K}. However, the best-fit compression factor of $\rm s\sim 360$ suggests an average density of approximately $\rm \sim 8000~ cm^{-3}$ in this region for dense molecular clouds.  Considering the observed density of dense molecular clouds as high as $\rm 10^5 ~ cm^{-3}$ \citep{2002ApJ...575..871R}, this correlation may be a posteriori as evidence that, for Cas A exhibiting double power-law features, the dominant GeV emission originates from re-accelerated particles in the background ``sea" of cosmic rays.

With an age of approximately 440 years, similar to that of Cas A, Tycho exhibits a gamma-ray spectrum that closely resembles that of Cas A in observations. As shown in Figure \ref{fig4}, using our model to interpret its observed characteristics, we note that the observational data can be effectively fitted. Around Tycho, the average gas density $\rm n_0 < 0.6~cm^{-3}$ was derived from the absence of thermal X-ray emission from the bright outer rim of the remnant \citep{2007ApJ...665..315C}. However, the fitting results of the observational data for Tycho's supernova remnant constrain the parameter s to $\sim 80$, indicating that the density of the surrounding molecular clouds is approximately $\rm 238~cm^{-3}$. This finding is consistent with previous research conclusions \citep{2011ApJ...729L..15T,2013MNRAS.429L..25Z,2016ApJ...826...34Z} which molecular gas surrounding Tycho may have a density as high as $\rm \sim 10^2~cm^{-3}$.
 
To extend our model, we applied it to SNR W44 and IC443.
Both can be reasonably explained by the shock re-acceleration and compression of pre-existing cosmic-ray particles. The corresponding parameters are listed in Table \ref{tab2}. It is important to note, however, that due to the lack of higher-energy data for both W44, we adopt a spectral index $ \alpha_2 = 4.0$ and a cutoff energy $\rm E_{br} = 1~TeV$ for W44 in this work.

Figure \ref{fig5} presents the fitting results about the SNRs W44. Unlike Cas A and Tycho, observations of W44 so far indicate that it energy spectra do not exhibit a distinct double power-law structure. This may be explained by scenarios in which the GeV–TeV $\gamma$-ray emission is dominated by $\pi^0$-decay resulting from distributions produced through shock re-acceleration and compression of pre-existing ambient CR particles. In such cases, the $\gamma$-ray spectrum is generally very soft.
As discussed in previous studies \citep{2012ApJ...761..133Y}, the high-energy gamma-ray emission from SNRs evolving in low-density media is dominated by inverse Compton scattering from accelerated electrons, while in high-density environments, the emission is governed by $\pi^{0}$ decay. Therefore, the double power-law structure is expected to be most prominent in SNRs located in moderately dense environments.

By the way, the gamma-ray spectrum observed by Fermi-LAT and HESS from the Galactic Center (GC) exhibits a clear double power-law shape. 
As discussed in our previous work \citep{2025arXiv250305261N}, the low-energy (GeV) emission may originate from interactions between trapped cosmic-ray particles from the background ``sea" and the dense gas environment near the Galactic center. In contrast, the bubble-like structure observed in the high-energy (TeV) regime is produced by protons accelerated during active phases of the Galactic center, via the same physical process.

It should be noted that nonlinear effects in shock acceleration can make the high-energy segment of the accelerated cosmic-ray particle spectrum flatter than expected in linear theory, while the low-energy segment becomes steeper \citep{2002APh....16..429B,2023ApJ...952..100N}. 
However, the sources considered in this work are located near dense molecular clouds. When the shock from the source sweeps through the region where dense molecular clouds are located, the shock slows down \citep{2010ApJ...724...59S,2016EPJWC.12104001G}. In the case of a slow shock, the spectrum of cosmic-ray particles re-accelerated by the shock almost retains the same spectral index as the injection spectrum (the spectrum of the background ``sea" CR particles) \citep{1983RPPh...46..973D}, 
so nonlinear effects are negligible in this scenario. When the shock sweeps through low-density regions, the accelerated spectrum is dominated by freely accelerated thermal particles, which constitute the high-energy component in the two-component structure. Therefore, nonlinear effects naturally do not affect the existence of the two-component structure. From an observational perspective, recent analysis of data from LHAASO on the extended emissions of IC443 suggests that the two-component structure may be a superposition of two origins \citep{2025arXiv251026112C}. Therefore, the likelihood of this structure being attributed to non-linear effects is relatively low.

In Figure \ref{fig6}, we also plot the re-accelerated proton spectrum and freshly accelerated electrons, which are obtained by the $\gamma$-ray observed data from their sources. 
The results presented in Figures \ref{fig3}, \ref{fig4} and \ref{fig5} demonstrate good agreement between our model predictions and observations. This not only supports the validity of our model in explaining the double power-law features observed in cosmic ray sources (particularly SNRs), but also lends further credence to our previously proposed spatially-dependent CR propagation framework. 
Our focus lies in highlighting the contribution of background ``sea" CRs to the observed gamma-ray from CR sources, particularly in environments where dense gas surrounds the source. We anticipate that future observations will identify more cosmic ray sources with double power-law features, enabling further tests of our model’s predictions and its general applicability.

\section{Conclusion and Summary} \label{sec:conclusion}
In this work, we propose a two-component model to explain the emission of gamma rays from several cosmic-ray (CR) sources. In our model, shock waves from the source reaccelerate the background ``sea" cosmic rays in the surrounding region, thereby introducing a new spectral component to the source. According to our previously proposed spatially dependent CR propagation framework \citep{2018PhRvD..97f3008G,2024PhRvD.109f3001Y,2024ApJ...974..276N,2025arXiv250305261N}, this component originates from distant sources and propagates through the Galaxy, making it softer than the component accelerated by the local source.

This model successfully explains some unique spectral features observed in cosmic ray sources, particularly the double power-law structure in SNR Cas A and IC443 newly revealed by LHAASO. 

Our model predicts that the double power-law spectral feature may be a universal characteristic of CR sources in dense environments, which strongly supports the spatially dependent CR propagation scenario: high-energy CRs are dominated by local sources, while low-energy CRs originate from the collective contribution of distant sources. We believe that LHAASO will identify more CR sources with double power-law features in the future, providing further validation for our model.


\acknowledgments

This work is supported in China by National Key $\rm R\&D$ Program of China under the grant 2024YFA1611402, and supported by the National Natural Science Foundation of China (12333006, 12275279, 12375103).


 \bibliographystyle{JHEP}
 \bibliography{biblio.bib}

@PREAMBLE{
 "\providecommand{\noopsort}[1]{}" 
 # "\providecommand{\singleletter}[1]{#1}%" 
}

@ARTICLE{2025ApJ...982L..33C,
       author = {{Cao}, Zhen and {Aharonian}, F. and {Bai}, Y.~X. and {Bao}, Y.~W. and {Bastieri}, D. and {Bi}, X.~J. and {Bi}, Y.~J. and {Bian}, W. and {Bukevich}, A.~V. and {Cai}, C.~M. and {Cao}, W.~Y. and {Cao}, Zhe and {Chang}, J. and {Chang}, J.~F. and {Chen}, A.~M. and {Chen}, E.~S. and {Chen}, H.~X. and {Chen}, Liang and {Chen}, Long and {Chen}, M.~J. and {Chen}, M.~L. and {Chen}, Q.~H. and {Chen}, S. and {Chen}, S.~H. and {Chen}, S.~Z. and {Chen}, T.~L. and {Chen}, X.~B. and {Chen}, X.~J. and {Chen}, Y. and {Cheng}, N. and {Cheng}, Y.~D. and {Chu}, M.~C. and {Cui}, M.~Y. and {Cui}, S.~W. and {Cui}, X.~H. and {Cui}, Y.~D. and {Dai}, B.~Z. and {Dai}, H.~L. and {Dai}, Z.~G. and {Danzengluobu} and {Diao}, Y.~X. and {Dong}, X.~Q. and {Duan}, K.~K. and {Fan}, J.~H. and {Fan}, Y.~Z. and {Fang}, J. and {Fang}, J.~H. and {Fang}, K. and {Feng}, C.~F. and {Feng}, H. and {Feng}, L. and {Feng}, S.~H. and {Feng}, X.~T. and {Feng}, Y. and {Feng}, Y.~L. and {Gabici}, S. and {Gao}, B. and {Gao}, C.~D. and {Gao}, Q. and {Gao}, W. and {Gao}, W.~K. and {Ge}, M.~M. and {Ge}, T.~T. and {Geng}, L.~S. and {Giacinti}, G. and {Gong}, G.~H. and {Gou}, Q.~B. and {Gu}, M.~H. and {Guo}, F.~L. and {Guo}, J. and {Guo}, X.~L. and {Guo}, Y.~Q. and {Guo}, Y.~Y. and {Han}, Y.~A. and {Hannuksela}, O.~A. and {Hasan}, M. and {He}, H.~H. and {He}, H.~N. and {He}, J.~Y. and {He}, X.~Y. and {He}, Y. and {Hernandez-Cadena}, S. and {Hor}, Y.~K. and {Hou}, B.~W. and {Hou}, C. and {Hou}, X. and {Hu}, H.~B. and {Hu}, S.~C. and {Huang}, C. and {Huang}, D.~H. and {Huang}, J.~J. and {Huang}, T.~Q. and {Huang}, W.~J. and {Huang}, X.~T. and {Huang}, X.~Y. and {Huang}, Y. and {Huang}, Y.~Y. and {Ji}, X.~L. and {Jia}, H.~Y. and {Jia}, K. and {Jiang}, H.~B. and {Jiang}, K. and {Jiang}, X.~W. and {Jiang}, Z.~J. and {Jin}, M. and {Kaci}, S. and {Kang}, M.~M. and {Karpikov}, I. and {Khangulyan}, D. and {Kuleshov}, D. and {Kurinov}, K. and {Li}, B.~B. and {Li}, Cheng and {Li}, Cong and {Li}, D. and {Li}, F. and {Li}, H.~B. and {Li}, H.~C. and {Li}, Jian and {Li}, Jie and {Li}, K. and {Li}, L. and {Li}, R.~L. and {Li}, S.~D. and {Li}, T.~Y. and {Li}, W.~L. and {Li}, X.~R. and {Li}, Xin and {Li}, Y.~Z. and {Li}, Zhe and {Li}, Zhuo and {Liang}, E.~W. and {Liang}, Y.~F. and {Lin}, S.~J. and {Liu}, B. and {Liu}, C. and {Liu}, D. and {Liu}, D.~B. and {Liu}, H. and {Liu}, H.~D. and {Liu}, J. and {Liu}, J.~L. and {Liu}, J.~R. and {Liu}, M.~Y. and {Liu}, R.~Y. and {Liu}, S.~M. and {Liu}, W. and {Liu}, X. and {Liu}, Y. and {Liu}, Y.~N. and {Lou}, Y.~Q. and {Luo}, Q. and {Luo}, Y. and {Lv}, H.~K. and {Ma}, B.~Q. and {Ma}, L.~L. and {Ma}, X.~H. and {Mao}, J.~R. and {Min}, Z. and {Mitthumsiri}, W. and {Mou}, G.~B. and {Mu}, H.~J. and {Nan}, Y.~C. and {Neronov}, A. and {Ng}, K.~C.~Y. and {Ni}, M.~Y. and {Nie}, L. and {Ou}, L.~J. and {Pattarakijwanich}, P. and {Pei}, Z.~Y. and {Qi}, J.~C. and {Qi}, M.~Y. and {Qin}, J.~J. and {Raza}, A. and {Ren}, C.~Y. and {Ruffolo}, D. and {S{\'a}iz}, A. and {Saeed}, M. and {Semikoz}, D. and {Shao}, L. and {Shchegolev}, O. and {Shen}, Y.~Z. and {Sheng}, X.~D. and {Shi}, Z.~D. and {Shu}, F.~W. and {Song}, H.~C. and {Stenkin}, Yu. V. and {Stepanov}, V. and {Su}, Y. and {Sun}, D.~X. and {Sun}, H. and {Sun}, Q.~N. and {Sun}, X.~N. and {Sun}, Z.~B. and {Tabasam}, N.~H. and {Takata}, J. and {Tam}, P.~H.~T. and {Tan}, H.~B. and {Tang}, Q.~W. and {Tang}, R.},
        title = "{Broadband {\ensuremath{\gamma}}-Ray Spectrum of Supernova Remnant Cassiopeia A}",
      journal = {apjl},
         year = 2025,
        month = mar,
       volume = {982},
       number = {1},
          eid = {L33},
        pages = {L33},
          doi = {10.3847/2041-8213/adb97c},
archivePrefix = {arXiv},
       eprint = {2502.04848},
 primaryClass = {astro-ph.HE},
       adsurl = {https://ui.adsabs.harvard.edu/abs/2025ApJ...982L..33C}
}

@ARTICLE{2018AdSpR..62.2845S,
       author = {{Sinitsyna}, Vera G. and {Sinitsyna}, Vera Y. and {Borisov}, Sergei S. and {Ivanov}, Ivan A. and {Klimov}, Anatolii I. and {Mirzafatikhov}, Rim M. and {Moseiko}, Nikolai I.},
        title = "{Shell-type supernova remnants as sources of cosmic rays}",
      journal = {Advances in Space Research},
         year = 2018,
        month = nov,
       volume = {62},
       number = {10},
        pages = {2845-2858},
          doi = {10.1016/j.asr.2017.04.007},
       adsurl = {https://ui.adsabs.harvard.edu/abs/2018AdSpR..62.2845S}
}

@ARTICLE{2022ApJS..260...53A,
       author = {{Abdollahi}, S. and {Acero}, F. and {Baldini}, L. and {Ballet}, J. and {Bastieri}, D. and {Bellazzini}, R. and {Berenji}, B. and {Berretta}, A. and {Bissaldi}, E. and {Blandford}, R.~D. and {Bloom}, E. and {Bonino}, R. and {Brill}, A. and {Britto}, R.~J. and {Bruel}, P. and {Burnett}, T.~H. and {Buson}, S. and {Cameron}, R.~A. and {Caputo}, R. and {Caraveo}, P.~A. and {Castro}, D. and {Chaty}, S. and {Cheung}, C.~C. and {Chiaro}, G. and {Cibrario}, N. and {Ciprini}, S. and {Coronado-Bl{\'a}zquez}, J. and {Crnogorcevic}, M. and {Cutini}, S. and {D'Ammando}, F. and {De Gaetano}, S. and {Digel}, S.~W. and {Di Lalla}, N. and {Dirirsa}, F. and {Di Venere}, L. and {Dom{\'\i}nguez}, A. and {Fallah Ramazani}, V. and {Fegan}, S.~J. and {Ferrara}, E.~C. and {Fiori}, A. and {Fleischhack}, H. and {Franckowiak}, A. and {Fukazawa}, Y. and {Funk}, S. and {Fusco}, P. and {Galanti}, G. and {Gammaldi}, V. and {Gargano}, F. and {Garrappa}, S. and {Gasparrini}, D. and {Giacchino}, F. and {Giglietto}, N. and {Giordano}, F. and {Giroletti}, M. and {Glanzman}, T. and {Green}, D. and {Grenier}, I.~A. and {Grondin}, M. -H. and {Guillemot}, L. and {Guiriec}, S. and {Gustafsson}, M. and {Harding}, A.~K. and {Hays}, E. and {Hewitt}, J.~W. and {Horan}, D. and {Hou}, X. and {J{\'o}hannesson}, G. and {Karwin}, C. and {Kayanoki}, T. and {Kerr}, M. and {Kuss}, M. and {Landriu}, D. and {Larsson}, S. and {Latronico}, L. and {Lemoine-Goumard}, M. and {Li}, J. and {Liodakis}, I. and {Longo}, F. and {Loparco}, F. and {Lott}, B. and {Lubrano}, P. and {Maldera}, S. and {Malyshev}, D. and {Manfreda}, A. and {Mart{\'\i}-Devesa}, G. and {Mazziotta}, M.~N. and {Mereu}, I. and {Meyer}, M. and {Michelson}, P.~F. and {Mirabal}, N. and {Mitthumsiri}, W. and {Mizuno}, T. and {Moiseev}, A.~A. and {Monzani}, M.~E. and {Morselli}, A. and {Moskalenko}, I.~V. and {Negro}, M. and {Nuss}, E. and {Omodei}, N. and {Orienti}, M. and {Orlando}, E. and {Paneque}, D. and {Pei}, Z. and {Perkins}, J.~S. and {Persic}, M. and {Pesce-Rollins}, M. and {Petrosian}, V. and {Pillera}, R. and {Poon}, H. and {Porter}, T.~A. and {Principe}, G. and {Rain{\`o}}, S. and {Rando}, R. and {Rani}, B. and {Razzano}, M. and {Razzaque}, S. and {Reimer}, A. and {Reimer}, O. and {Reposeur}, T. and {S{\'a}nchez-Conde}, M. and {Saz Parkinson}, P.~M. and {Scotton}, L. and {Serini}, D. and {Sgr{\`o}}, C. and {Siskind}, E.~J. and {Smith}, D.~A. and {Spandre}, G. and {Spinelli}, P. and {Sueoka}, K. and {Suson}, D.~J. and {Tajima}, H. and {Tak}, D. and {Thayer}, J.~B. and {Thompson}, D.~J. and {Torres}, D.~F. and {Troja}, E. and {Valverde}, J. and {Wood}, K. and {Zaharijas}, G.},
        title = "{Incremental Fermi Large Area Telescope Fourth Source Catalog}",
      journal = {apjs},
         year = 2022,
        month = jun,
       volume = {260},
       number = {2},
          eid = {53},
        pages = {53},
          doi = {10.3847/1538-4365/ac6751},
archivePrefix = {arXiv},
       eprint = {2201.11184},
 primaryClass = {astro-ph.HE},
       adsurl = {https://ui.adsabs.harvard.edu/abs/2022ApJS..260...53A}
}

@ARTICLE{2021MNRAS.502..472A,
       author = {{Araya}, M. and {Herrera}, C.},
        title = "{A source of gamma rays coincident with the shell of the supernova remnant CTB 80}",
      journal = {mnras},
         year = 2021,
        month = mar,
       volume = {502},
       number = {1},
        pages = {472-477},
          doi = {10.1093/mnras/stab101},
archivePrefix = {arXiv},
       eprint = {2101.04020},
 primaryClass = {astro-ph.HE},
       adsurl = {https://ui.adsabs.harvard.edu/abs/2021MNRAS.502..472A}
}

@ARTICLE{2013Sci...339..807A,
       author = {{Ackermann}, M. and {Ajello}, M. and {Allafort}, A. and {Baldini}, L. and {Ballet}, J. and {Barbiellini}, G. and {Baring}, M.~G. and {Bastieri}, D. and {Bechtol}, K. and {Bellazzini}, R. and {Blandford}, R.~D. and {Bloom}, E.~D. and {Bonamente}, E. and {Borgland}, A.~W. and {Bottacini}, E. and {Brandt}, T.~J. and {Bregeon}, J. and {Brigida}, M. and {Bruel}, P. and {Buehler}, R. and {Busetto}, G. and {Buson}, S. and {Caliandro}, G.~A. and {Cameron}, R.~A. and {Caraveo}, P.~A. and {Casandjian}, J.~M. and {Cecchi}, C. and {{\c{C}}elik}, {\"O}. and {Charles}, E. and {Chaty}, S. and {Chaves}, R.~C.~G. and {Chekhtman}, A. and {Cheung}, C.~C. and {Chiang}, J. and {Chiaro}, G. and {Cillis}, A.~N. and {Ciprini}, S. and {Claus}, R. and {Cohen-Tanugi}, J. and {Cominsky}, L.~R. and {Conrad}, J. and {Corbel}, S. and {Cutini}, S. and {D'Ammando}, F. and {de Angelis}, A. and {de Palma}, F. and {Dermer}, C.~D. and {do Couto e Silva}, E. and {Drell}, P.~S. and {Drlica-Wagner}, A. and {Falletti}, L. and {Favuzzi}, C. and {Ferrara}, E.~C. and {Franckowiak}, A. and {Fukazawa}, Y. and {Funk}, S. and {Fusco}, P. and {Gargano}, F. and {Germani}, S. and {Giglietto}, N. and {Giommi}, P. and {Giordano}, F. and {Giroletti}, M. and {Glanzman}, T. and {Godfrey}, G. and {Grenier}, I.~A. and {Grondin}, M. -H. and {Grove}, J.~E. and {Guiriec}, S. and {Hadasch}, D. and {Hanabata}, Y. and {Harding}, A.~K. and {Hayashida}, M. and {Hayashi}, K. and {Hays}, E. and {Hewitt}, J.~W. and {Hill}, A.~B. and {Hughes}, R.~E. and {Jackson}, M.~S. and {Jogler}, T. and {J{\'o}hannesson}, G. and {Johnson}, A.~S. and {Kamae}, T. and {Kataoka}, J. and {Katsuta}, J. and {Kn{\"o}dlseder}, J. and {Kuss}, M. and {Lande}, J. and {Larsson}, S. and {Latronico}, L. and {Lemoine-Goumard}, M. and {Longo}, F. and {Loparco}, F. and {Lovellette}, M.~N. and {Lubrano}, P. and {Madejski}, G.~M. and {Massaro}, F. and {Mayer}, M. and {Mazziotta}, M.~N. and {McEnery}, J.~E. and {Mehault}, J. and {Michelson}, P.~F. and {Mignani}, R.~P. and {Mitthumsiri}, W. and {Mizuno}, T. and {Moiseev}, A.~A. and {Monzani}, M.~E. and {Morselli}, A. and {Moskalenko}, I.~V. and {Murgia}, S. and {Nakamori}, T. and {Nemmen}, R. and {Nuss}, E. and {Ohno}, M. and {Ohsugi}, T. and {Omodei}, N. and {Orienti}, M. and {Orlando}, E. and {Ormes}, J.~F. and {Paneque}, D. and {Perkins}, J.~S. and {Pesce-Rollins}, M. and {Piron}, F. and {Pivato}, G. and {Rain{\`o}}, S. and {Rando}, R. and {Razzano}, M. and {Razzaque}, S. and {Reimer}, A. and {Reimer}, O. and {Ritz}, S. and {Romoli}, C. and {S{\'a}nchez-Conde}, M. and {Schulz}, A. and {Sgr{\`o}}, C. and {Simeon}, P.~E. and {Siskind}, E.~J. and {Smith}, D.~A. and {Spandre}, G. and {Spinelli}, P. and {Stecker}, F.~W. and {Strong}, A.~W. and {Suson}, D.~J. and {Tajima}, H. and {Takahashi}, H. and {Takahashi}, T. and {Tanaka}, T. and {Thayer}, J.~G. and {Thayer}, J.~B. and {Thompson}, D.~J. and {Thorsett}, S.~E. and {Tibaldo}, L. and {Tibolla}, O. and {Tinivella}, M. and {Troja}, E. and {Uchiyama}, Y. and {Usher}, T.~L. and {Vandenbroucke}, J. and {Vasileiou}, V. and {Vianello}, G. and {Vitale}, V. and {Waite}, A.~P. and {Werner}, M. and {Winer}, B.~L. and {Wood}, K.~S. and {Wood}, M. and {Yamazaki}, R. and {Yang}, Z. and {Zimmer}, S.},
        title = "{Detection of the Characteristic Pion-Decay Signature in Supernova Remnants}",
      journal = {Science},
         year = 2013,
        month = feb,
       volume = {339},
       number = {6121},
        pages = {807-811},
          doi = {10.1126/science.1231160},
archivePrefix = {arXiv},
       eprint = {1302.3307},
 primaryClass = {astro-ph.HE},
       adsurl = {https://ui.adsabs.harvard.edu/abs/2013Sci...339..807A}
}

@ARTICLE{2015PhRvL.114q1103A,
       author = {{Aguilar}, M. and {Aisa}, D. and {Alpat}, B. and {Alvino}, A. and {Ambrosi}, G. and {Andeen}, K. and {Arruda}, L. and {Attig}, N. and {Azzarello}, P. and {Bachlechner}, A. and {Barao}, F. and {Barrau}, A. and {Barrin}, L. and {Bartoloni}, A. and {Basara}, L. and {Battarbee}, M. and {Battiston}, R. and {Bazo}, J. and {Becker}, U. and {Behlmann}, M. and {Beischer}, B. and {Berdugo}, J. and {Bertucci}, B. and {Bigongiari}, G. and {Bindi}, V. and {Bizzaglia}, S. and {Bizzarri}, M. and {Boella}, G. and {de Boer}, W. and {Bollweg}, K. and {Bonnivard}, V. and {Borgia}, B. and {Borsini}, S. and {Boschini}, M.~J. and {Bourquin}, M. and {Burger}, J. and {Cadoux}, F. and {Cai}, X.~D. and {Capell}, M. and {Caroff}, S. and {Casaus}, J. and {Cascioli}, V. and {Castellini}, G. and {Cernuda}, I. and {Cerreta}, D. and {Cervelli}, F. and {Chae}, M.~J. and {Chang}, Y.~H. and {Chen}, A.~I. and {Chen}, H. and {Cheng}, G.~M. and {Chen}, H.~S. and {Cheng}, L. and {Chou}, H.~Y. and {Choumilov}, E. and {Choutko}, V. and {Chung}, C.~H. and {Clark}, C. and {Clavero}, R. and {Coignet}, G. and {Consolandi}, C. and {Contin}, A. and {Corti}, C. and {Gil}, E. Cortina and {Coste}, B. and {Creus}, W. and {Crispoltoni}, M. and {Cui}, Z. and {Dai}, Y.~M. and {Delgado}, C. and {Della Torre}, S. and {Demirk{\"o}z}, M.~B. and {Derome}, L. and {Di Falco}, S. and {Di Masso}, L. and {Dimiccoli}, F. and {D{\'\i}az}, C. and {von Doetinchem}, P. and {Donnini}, F. and {Du}, W.~J. and {Duranti}, M. and {D'Urso}, D. and {Eline}, A. and {Eppling}, F.~J. and {Eronen}, T. and {Fan}, Y.~Y. and {Farnesini}, L. and {Feng}, J. and {Fiandrini}, E. and {Fiasson}, A. and {Finch}, E. and {Fisher}, P. and {Galaktionov}, Y. and {Gallucci}, G. and {Garc{\'\i}a}, B. and {Garc{\'\i}a-L{\'o}pez}, R. and {Gargiulo}, C. and {Gast}, H. and {Gebauer}, I. and {Gervasi}, M. and {Ghelfi}, A. and {Gillard}, W. and {Giovacchini}, F. and {Goglov}, P. and {Gong}, J. and {Goy}, C. and {Grabski}, V. and {Grandi}, D. and {Graziani}, M. and {Guandalini}, C. and {Guerri}, I. and {Guo}, K.~H. and {Haas}, D. and {Habiby}, M. and {Haino}, S. and {Han}, K.~C. and {He}, Z.~H. and {Heil}, M. and {Hoffman}, J. and {Hsieh}, T.~H. and {Huang}, Z.~C. and {Huh}, C. and {Incagli}, M. and {Ionica}, M. and {Jang}, W.~Y. and {Jinchi}, H. and {Kanishev}, K. and {Kim}, G.~N. and {Kim}, K.~S. and {Kirn}, Th. and {Kossakowski}, R. and {Kounina}, O. and {Kounine}, A. and {Koutsenko}, V. and {Krafczyk}, M.~S. and {La Vacca}, G. and {Laudi}, E. and {Laurenti}, G. and {Lazzizzera}, I. and {Lebedev}, A. and {Lee}, H.~T. and {Lee}, S.~C. and {Leluc}, C. and {Levi}, G. and {Li}, H.~L. and {Li}, J.~Q. and {Li}, Q. and {Li}, Q. and {Li}, T.~X. and {Li}, W. and {Li}, Y. and {Li}, Z.~H. and {Li}, Z.~Y. and {Lim}, S. and {Lin}, C.~H. and {Lipari}, P. and {Lippert}, T. and {Liu}, D. and {Liu}, H. and {Lolli}, M. and {Lomtadze}, T. and {Lu}, M.~J. and {Lu}, S.~Q. and {Lu}, Y.~S. and {Luebelsmeyer}, K. and {Luo}, J.~Z. and {Lv}, S.~S. and {Majka}, R. and {Ma{\~n}{\'a}}, C. and {Mar{\'\i}n}, J. and {Martin}, T. and {Mart{\'\i}nez}, G. and {Masi}, N. and {Maurin}, D. and {Menchaca-Rocha}, A. and {Meng}, Q. and {Mo}, D.~C. and {Morescalchi}, L. and {Mott}, P. and {M{\"u}ller}, M. and {Ni}, J.~Q. and {Nikonov}, N. and {Nozzoli}, F. and {Nunes}, P. and {Obermeier}, A. and {Oliva}, A. and {Orcinha}, M. and {Palmonari}, F. and {Palomares}, C. and {Paniccia}, M. and {Papi}, A. and {Pauluzzi}, M. and {Pedreschi}, E. and {Pensotti}, S. and {Pereira}, R. and {Picot-Clemente}, N. and {Pilo}, F. and {Piluso}, A. and {Pizzolotto}, C. and {Plyaskin}, V.},
        title = "{Precision Measurement of the Proton Flux in Primary Cosmic Rays from Rigidity 1 GV to 1.8 TV with the Alpha Magnetic Spectrometer on the International Space Station}",
      journal = {prl},
         year = 2015,
        month = may,
       volume = {114},
       number = {17},
          eid = {171103},
        pages = {171103},
          doi = {10.1103/PhysRevLett.114.171103},
       adsurl = {https://ui.adsabs.harvard.edu/abs/2015PhRvL.114q1103A}
}

@ARTICLE{2015PhRvL.115u1101A,
       author = {{Aguilar}, M. and {Aisa}, D. and {Alpat}, B. and {Alvino}, A. and {Ambrosi}, G. and {Andeen}, K. and {Arruda}, L. and {Attig}, N. and {Azzarello}, P. and {Bachlechner}, A. and {Barao}, F. and {Barrau}, A. and {Barrin}, L. and {Bartoloni}, A. and {Basara}, L. and {Battarbee}, M. and {Battiston}, R. and {Bazo}, J. and {Becker}, U. and {Behlmann}, M. and {Beischer}, B. and {Berdugo}, J. and {Bertucci}, B. and {Bindi}, V. and {Bizzaglia}, S. and {Bizzarri}, M. and {Boella}, G. and {de Boer}, W. and {Bollweg}, K. and {Bonnivard}, V. and {Borgia}, B. and {Borsini}, S. and {Boschini}, M.~J. and {Bourquin}, M. and {Burger}, J. and {Cadoux}, F. and {Cai}, X.~D. and {Capell}, M. and {Caroff}, S. and {Casaus}, J. and {Castellini}, G. and {Cernuda}, I. and {Cerreta}, D. and {Cervelli}, F. and {Chae}, M.~J. and {Chang}, Y.~H. and {Chen}, A.~I. and {Chen}, G.~M. and {Chen}, H. and {Chen}, H.~S. and {Cheng}, L. and {Chou}, H.~Y. and {Choumilov}, E. and {Choutko}, V. and {Chung}, C.~H. and {Clark}, C. and {Clavero}, R. and {Coignet}, G. and {Consolandi}, C. and {Contin}, A. and {Corti}, C. and {Gil}, E. Cortina and {Coste}, B. and {Creus}, W. and {Crispoltoni}, M. and {Cui}, Z. and {Dai}, Y.~M. and {Delgado}, C. and {Della Torre}, S. and {Demirk{\"o}z}, M.~B. and {Derome}, L. and {Di Falco}, S. and {Di Masso}, L. and {Dimiccoli}, F. and {D{\'\i}az}, C. and {von Doetinchem}, P. and {Donnini}, F. and {Duranti}, M. and {D'Urso}, D. and {Egorov}, A. and {Eline}, A. and {Eppling}, F.~J. and {Eronen}, T. and {Fan}, Y.~Y. and {Farnesini}, L. and {Feng}, J. and {Fiandrini}, E. and {Fiasson}, A. and {Finch}, E. and {Fisher}, P. and {Formato}, V. and {Galaktionov}, Y. and {Gallucci}, G. and {Garc{\'\i}a}, B. and {Garc{\'\i}a-L{\'o}pez}, R. and {Gargiulo}, C. and {Gast}, H. and {Gebauer}, I. and {Gervasi}, M. and {Ghelfi}, A. and {Giovacchini}, F. and {Goglov}, P. and {Gong}, J. and {Goy}, C. and {Grabski}, V. and {Grandi}, D. and {Graziani}, M. and {Guandalini}, C. and {Guerri}, I. and {Guo}, K.~H. and {Haas}, D. and {Habiby}, M. and {Haino}, S. and {Han}, K.~C. and {He}, Z.~H. and {Heil}, M. and {Hoffman}, J. and {Hsieh}, T.~H. and {Huang}, Z.~C. and {Huh}, C. and {Incagli}, M. and {Ionica}, M. and {Jang}, W.~Y. and {Jinchi}, H. and {Kanishev}, K. and {Kim}, G.~N. and {Kim}, K.~S. and {Kirn}, Th. and {Korkmaz}, M.~A. and {Kossakowski}, R. and {Kounina}, O. and {Kounine}, A. and {Koutsenko}, V. and {Krafczyk}, M.~S. and {La Vacca}, G. and {Laudi}, E. and {Laurenti}, G. and {Lazzizzera}, I. and {Lebedev}, A. and {Lee}, H.~T. and {Lee}, S.~C. and {Leluc}, C. and {Li}, H.~L. and {Li}, J.~Q. and {Li}, J.~Q. and {Li}, Q. and {Li}, Q. and {Li}, T.~X. and {Li}, W. and {Li}, Y. and {Li}, Z.~H. and {Li}, Z.~Y. and {Lim}, S. and {Lin}, C.~H. and {Lipari}, P. and {Lippert}, T. and {Liu}, D. and {Liu}, H. and {Liu}, Hu and {Lolli}, M. and {Lomtadze}, T. and {Lu}, M.~J. and {Lu}, S.~Q. and {Lu}, Y.~S. and {Luebelsmeyer}, K. and {Luo}, F. and {Luo}, J.~Z. and {Lv}, S.~S. and {Majka}, R. and {Ma{\~n}{\'a}}, C. and {Mar{\'\i}n}, J. and {Martin}, T. and {Mart{\'\i}nez}, G. and {Masi}, N. and {Maurin}, D. and {Menchaca-Rocha}, A. and {Meng}, Q. and {Mo}, D.~C. and {Morescalchi}, L. and {Mott}, P. and {M{\"u}ller}, M. and {Nelson}, T. and {Ni}, J.~Q. and {Nikonov}, N. and {Nozzoli}, F. and {Nunes}, P. and {Obermeier}, A. and {Oliva}, A. and {Orcinha}, M. and {Palmonari}, F. and {Palomares}, C. and {Paniccia}, M. and {Papi}, A. and {Pauluzzi}, M. and {Pedreschi}, E. and {Pensotti}, S. and {Pereira}, R. and {Picot-Clemente}, N. and {Pilo}, F. and {Piluso}, A.},
        title = "{Precision Measurement of the Helium Flux in Primary Cosmic Rays of Rigidities 1.9 GV to 3 TV with the Alpha Magnetic Spectrometer on the International Space Station}",
      journal = {prl},
         year = 2015,
        month = nov,
       volume = {115},
       number = {21},
          eid = {211101},
        pages = {211101},
          doi = {10.1103/PhysRevLett.115.211101},
       adsurl = {https://ui.adsabs.harvard.edu/abs/2015PhRvL.115u1101A}
}

@ARTICLE{2017PhRvL.119y1101A,
       author = {{Aguilar}, M. and {Ali Cavasonza}, L. and {Alpat}, B. and {Ambrosi}, G. and {Arruda}, L. and {Attig}, N. and {Aupetit}, S. and {Azzarello}, P. and {Bachlechner}, A. and {Barao}, F. and {Barrau}, A. and {Barrin}, L. and {Bartoloni}, A. and {Basara}, L. and {Ba{\c{s}}e{\v{g}}mez-du Pree}, S. and {Battarbee}, M. and {Battiston}, R. and {Becker}, U. and {Behlmann}, M. and {Beischer}, B. and {Berdugo}, J. and {Bertucci}, B. and {Bindel}, K.~F. and {Bindi}, V. and {de Boer}, W. and {Bollweg}, K. and {Bonnivard}, V. and {Borgia}, B. and {Boschini}, M.~J. and {Bourquin}, M. and {Bueno}, E.~F. and {Burger}, J. and {Burger}, W.~J. and {Cadoux}, F. and {Cai}, X.~D. and {Capell}, M. and {Caroff}, S. and {Casaus}, J. and {Castellini}, G. and {Cervelli}, F. and {Chae}, M.~J. and {Chang}, Y.~H. and {Chen}, A.~I. and {Chen}, G.~M. and {Chen}, H.~S. and {Cheng}, L. and {Chou}, H.~Y. and {Choumilov}, E. and {Choutko}, V. and {Chung}, C.~H. and {Clark}, C. and {Clavero}, R. and {Coignet}, G. and {Consolandi}, C. and {Contin}, A. and {Corti}, C. and {Creus}, W. and {Crispoltoni}, M. and {Cui}, Z. and {Dadzie}, K. and {Dai}, Y.~M. and {Datta}, A. and {Delgado}, C. and {Della Torre}, S. and {Demakov}, O. and {Demirk{\"o}z}, M.~B. and {Derome}, L. and {Di Falco}, S. and {Dimiccoli}, F. and {D{\'\i}az}, C. and {von Doetinchem}, P. and {Dong}, F. and {Donnini}, F. and {Duranti}, M. and {D'Urso}, D. and {Egorov}, A. and {Eline}, A. and {Eronen}, T. and {Feng}, J. and {Fiandrini}, E. and {Fisher}, P. and {Formato}, V. and {Galaktionov}, Y. and {Gallucci}, G. and {Garc{\'\i}a-L{\'o}pez}, R.~J. and {Gargiulo}, C. and {Gast}, H. and {Gebauer}, I. and {Gervasi}, M. and {Ghelfi}, A. and {Giovacchini}, F. and {G{\'o}mez-Coral}, D.~M. and {Gong}, J. and {Goy}, C. and {Grabski}, V. and {Grandi}, D. and {Graziani}, M. and {Guo}, K.~H. and {Haino}, S. and {Han}, K.~C. and {He}, Z.~H. and {Heil}, M. and {Hoffman}, J. and {Hsieh}, T.~H. and {Huang}, H. and {Huang}, Z.~C. and {Huh}, C. and {Incagli}, M. and {Ionica}, M. and {Jang}, W.~Y. and {Jia}, Yi and {Jinchi}, H. and {Kang}, S.~C. and {Kanishev}, K. and {Khiali}, B. and {Kim}, G.~N. and {Kim}, K.~S. and {Kirn}, Th. and {Konak}, C. and {Kounina}, O. and {Kounine}, A. and {Koutsenko}, V. and {Kulemzin}, A. and {La Vacca}, G. and {Laudi}, E. and {Laurenti}, G. and {Lazzizzera}, I. and {Lebedev}, A. and {Lee}, H.~T. and {Lee}, S.~C. and {Leluc}, C. and {Li}, H.~S. and {Li}, J.~Q. and {Li}, Q. and {Li}, T.~X. and {Li}, Y. and {Li}, Z.~H. and {Li}, Z.~Y. and {Lim}, S. and {Lin}, C.~H. and {Lipari}, P. and {Lippert}, T. and {Liu}, D. and {Liu}, Hu and {Lordello}, V.~D. and {Lu}, S.~Q. and {Lu}, Y.~S. and {Luebelsmeyer}, K. and {Luo}, F. and {Luo}, J.~Z. and {Lyu}, S.~S. and {Machate}, F. and {Ma{\~n}{\'a}}, C. and {Mar{\'\i}n}, J. and {Martin}, T. and {Mart{\'\i}nez}, G. and {Masi}, N. and {Maurin}, D. and {Menchaca-Rocha}, A. and {Meng}, Q. and {Mikuni}, V.~M. and {Mo}, D.~C. and {Mott}, P. and {Nelson}, T. and {Ni}, J.~Q. and {Nikonov}, N. and {Nozzoli}, F. and {Oliva}, A. and {Orcinha}, M. and {Palmonari}, F. and {Palomares}, C. and {Paniccia}, M. and {Pauluzzi}, M. and {Pensotti}, S. and {Perrina}, C. and {Phan}, H.~D. and {Picot-Clemente}, N. and {Pilo}, F. and {Pizzolotto}, C. and {Plyaskin}, V. and {Pohl}, M. and {Poireau}, V. and {Quadrani}, L. and {Qi}, X.~M. and {Qin}, X. and {Qu}, Z.~Y. and {R{\"a}ih{\"a}}, T. and {Rancoita}, P.~G. and {Rapin}, D. and {Ricol}, J.~S. and {Rosier-Lees}, S. and {Rozhkov}, A. and {Rozza}, D. and {Sagdeev}, R. and {Schael}, S. and {Schmidt}, S.~M. and {Schulz von Dratzig}, A. and {Schwering}, G. and {Seo}, E.~S. and {Shan}, B.~S.},
        title = "{Observation of the Identical Rigidity Dependence of He, C, and O Cosmic Rays at High Rigidities by the Alpha Magnetic Spectrometer on the International Space Station}",
      journal = {prl},
         year = 2017,
        month = dec,
       volume = {119},
       number = {25},
          eid = {251101},
        pages = {251101},
          doi = {10.1103/PhysRevLett.119.251101},
       adsurl = {https://ui.adsabs.harvard.edu/abs/2017PhRvL.119y1101A}
}

@ARTICLE{2018PhRvL.120b1101A,
       author = {{Aguilar}, M. and {Ali Cavasonza}, L. and {Ambrosi}, G. and {Arruda}, L. and {Attig}, N. and {Aupetit}, S. and {Azzarello}, P. and {Bachlechner}, A. and {Barao}, F. and {Barrau}, A. and {Barrin}, L. and {Bartoloni}, A. and {Basara}, L. and {Ba{\c{s}}e{\v{g}}mez-du Pree}, S. and {Battarbee}, M. and {Battiston}, R. and {Becker}, U. and {Behlmann}, M. and {Beischer}, B. and {Berdugo}, J. and {Bertucci}, B. and {Bindel}, K.~F. and {Bindi}, V. and {de Boer}, W. and {Bollweg}, K. and {Bonnivard}, V. and {Borgia}, B. and {Boschini}, M.~J. and {Bourquin}, M. and {Bueno}, E.~F. and {Burger}, J. and {Burger}, W.~J. and {Cadoux}, F. and {Cai}, X.~D. and {Capell}, M. and {Caroff}, S. and {Casaus}, J. and {Castellini}, G. and {Cervelli}, F. and {Chae}, M.~J. and {Chang}, Y.~H. and {Chen}, A.~I. and {Chen}, G.~M. and {Chen}, H.~S. and {Cheng}, L. and {Chou}, H.~Y. and {Choumilov}, E. and {Choutko}, V. and {Chung}, C.~H. and {Clark}, C. and {Clavero}, R. and {Coignet}, G. and {Consolandi}, C. and {Contin}, A. and {Corti}, C. and {Creus}, W. and {Crispoltoni}, M. and {Cui}, Z. and {Dadzie}, K. and {Dai}, Y.~M. and {Datta}, A. and {Delgado}, C. and {Della Torre}, S. and {Demirk{\"o}z}, M.~B. and {Derome}, L. and {Di Falco}, S. and {Dimiccoli}, F. and {D{\'\i}az}, C. and {von Doetinchem}, P. and {Dong}, F. and {Donnini}, F. and {Duranti}, M. and {D'Urso}, D. and {Egorov}, A. and {Eline}, A. and {Eronen}, T. and {Feng}, J. and {Fiandrini}, E. and {Fisher}, P. and {Formato}, V. and {Galaktionov}, Y. and {Gallucci}, G. and {Garc{\'\i}a-L{\'o}pez}, R.~J. and {Gargiulo}, C. and {Gast}, H. and {Gebauer}, I. and {Gervasi}, M. and {Ghelfi}, A. and {Giovacchini}, F. and {G{\'o}mez-Coral}, D.~M. and {Gong}, J. and {Goy}, C. and {Grabski}, V. and {Grandi}, D. and {Graziani}, M. and {Guo}, K.~H. and {Haino}, S. and {Han}, K.~C. and {He}, Z.~H. and {Heil}, M. and {Hsieh}, T.~H. and {Huang}, H. and {Huang}, Z.~C. and {Huh}, C. and {Incagli}, M. and {Ionica}, M. and {Jang}, W.~Y. and {Jia}, Yi and {Jinchi}, H. and {Kang}, S.~C. and {Kanishev}, K. and {Khiali}, B. and {Kim}, G.~N. and {Kim}, K.~S. and {Kirn}, Th. and {Konak}, C. and {Kounina}, O. and {Kounine}, A. and {Koutsenko}, V. and {Kulemzin}, A. and {La Vacca}, G. and {Laudi}, E. and {Laurenti}, G. and {Lazzizzera}, I. and {Lebedev}, A. and {Lee}, H.~T. and {Lee}, S.~C. and {Leluc}, C. and {Li}, H.~S. and {Li}, J.~Q. and {Li}, Q. and {Li}, T.~X. and {Li}, Y. and {Li}, Z.~H. and {Li}, Z.~Y. and {Lim}, S. and {Lin}, C.~H. and {Lipari}, P. and {Lippert}, T. and {Liu}, D. and {Liu}, Hu and {Lordello}, V.~D. and {Lu}, S.~Q. and {Lu}, Y.~S. and {Luebelsmeyer}, K. and {Luo}, F. and {Luo}, J.~Z. and {Lyu}, S.~S. and {Machate}, F. and {Ma{\~n}{\'a}}, C. and {Mar{\'\i}n}, J. and {Martin}, T. and {Mart{\'\i}nez}, G. and {Masi}, N. and {Maurin}, D. and {Menchaca-Rocha}, A. and {Meng}, Q. and {Mikuni}, V.~M. and {Mo}, D.~C. and {Mott}, P. and {Nelson}, T. and {Ni}, J.~Q. and {Nikonov}, N. and {Nozzoli}, F. and {Oliva}, A. and {Orcinha}, M. and {Palermo}, M. and {Palmonari}, F. and {Palomares}, C. and {Paniccia}, M. and {Pauluzzi}, M. and {Pensotti}, S. and {Perrina}, C. and {Phan}, H.~D. and {Picot-Clemente}, N. and {Pilo}, F. and {Pizzolotto}, C. and {Plyaskin}, V. and {Pohl}, M. and {Poireau}, V. and {Quadrani}, L. and {Qi}, X.~M. and {Qin}, X. and {Qu}, Z.~Y. and {R{\"a}ih{\"a}}, T. and {Rancoita}, P.~G. and {Rapin}, D. and {Ricol}, J.~S. and {Rosier-Lees}, S. and {Rozhkov}, A. and {Rozza}, D. and {Sagdeev}, R. and {Schael}, S. and {Schmidt}, S.~M. and {Schulz von Dratzig}, A. and {Schwering}, G. and {Seo}, E.~S. and {Shan}, B.~S. and {Shi}, J.~Y. and {Siedenburg}, T.},
        title = "{Observation of New Properties of Secondary Cosmic Rays Lithium, Beryllium, and Boron by the Alpha Magnetic Spectrometer on the International Space Station}",
      journal = {prl},
         year = 2018,
        month = jan,
       volume = {120},
       number = {2},
          eid = {021101},
        pages = {021101},
          doi = {10.1103/PhysRevLett.120.021101},
       adsurl = {https://ui.adsabs.harvard.edu/abs/2018PhRvL.120b1101A}
}

@ARTICLE{2018PhRvL.121e1103A,
       author = {{Aguilar}, M. and {Ali Cavasonza}, L. and {Alpat}, B. and {Ambrosi}, G. and {Arruda}, L. and {Attig}, N. and {Aupetit}, S. and {Azzarello}, P. and {Bachlechner}, A. and {Barao}, F. and {Barrau}, A. and {Barrin}, L. and {Bartoloni}, A. and {Basara}, L. and {Ba{\c{s}}e{\v{g}}mez-du Pree}, S. and {Battarbee}, M. and {Battiston}, R. and {Becker}, U. and {Behlmann}, M. and {Beischer}, B. and {Berdugo}, J. and {Bertucci}, B. and {Bindel}, K.~F. and {Bindi}, V. and {de Boer}, W. and {Bollweg}, K. and {Bonnivard}, V. and {Borgia}, B. and {Boschini}, M.~J. and {Bourquin}, M. and {Bueno}, E.~F. and {Burger}, J. and {Burger}, W.~J. and {Cai}, X.~D. and {Capell}, M. and {Caroff}, S. and {Casaus}, J. and {Castellini}, G. and {Cervelli}, F. and {Chang}, Y.~H. and {Chen}, A.~I. and {Chen}, G.~M. and {Chen}, H.~S. and {Chen}, Y. and {Cheng}, L. and {Chou}, H.~Y. and {Choumilov}, E. and {Choutko}, V. and {Chung}, C.~H. and {Clark}, C. and {Clavero}, R. and {Coignet}, G. and {Consolandi}, C. and {Contin}, A. and {Corti}, C. and {Creus}, W. and {Crispoltoni}, M. and {Cui}, Z. and {Dadzie}, K. and {Dai}, Y.~M. and {Datta}, A. and {Delgado}, C. and {Della Torre}, S. and {Demirk{\"o}z}, M.~B. and {Derome}, L. and {Di Falco}, S. and {Dimiccoli}, F. and {D{\'\i}az}, C. and {von Doetinchem}, P. and {Dong}, F. and {Donnini}, F. and {Duranti}, M. and {Egorov}, A. and {Eline}, A. and {Eronen}, T. and {Feng}, J. and {Fiandrini}, E. and {Fisher}, P. and {Formato}, V. and {Galaktionov}, Y. and {Gallucci}, G. and {Garc{\'\i}a-L{\'o}pez}, R.~J. and {Gargiulo}, C. and {Gast}, H. and {Gebauer}, I. and {Gervasi}, M. and {Ghelfi}, A. and {Giovacchini}, F. and {G{\'o}mez-Coral}, D.~M. and {Gong}, J. and {Goy}, C. and {Grabski}, V. and {Grandi}, D. and {Graziani}, M. and {Guo}, K.~H. and {Haino}, S. and {Han}, K.~C. and {He}, Z.~H. and {Heil}, M. and {Hsieh}, T.~H. and {Huang}, H. and {Huang}, Z.~C. and {Incagli}, M. and {Jia}, Yi and {Jinchi}, H. and {Kanishev}, K. and {Khiali}, B. and {Kirn}, Th. and {Konak}, C. and {Kounina}, O. and {Kounine}, A. and {Koutsenko}, V. and {Kulemzin}, A. and {La Vacca}, G. and {Laudi}, E. and {Laurenti}, G. and {Lazzizzera}, I. and {Lebedev}, A. and {Lee}, H.~T. and {Lee}, S.~C. and {Leluc}, C. and {Li}, H.~S. and {Li}, J.~Q. and {Li}, Q. and {Li}, T.~X. and {Li}, Z.~H. and {Li}, Z.~Y. and {Lin}, C.~H. and {Lipari}, P. and {Lippert}, T. and {Liu}, D. and {Liu}, Hu and {Liu}, Z. and {Lordello}, V.~D. and {Lu}, S.~Q. and {Lu}, Y.~S. and {Luebelsmeyer}, K. and {Luo}, F. and {Luo}, J.~Z. and {Lyu}, S.~S. and {Machate}, F. and {Ma{\~n}{\'a}}, C. and {Mar{\'\i}n}, J. and {Martin}, T. and {Mart{\'\i}nez}, G. and {Masi}, N. and {Maurin}, D. and {Menchaca-Rocha}, A. and {Meng}, Q. and {Mikuni}, V.~M. and {Mo}, D.~C. and {Mott}, P. and {Mussolin}, L. and {Nelson}, T. and {Ni}, J.~Q. and {Nikonov}, N. and {Nozzoli}, F. and {Oliva}, A. and {Orcinha}, M. and {Palermo}, M. and {Palmonari}, F. and {Palomares}, C. and {Paniccia}, M. and {Pauluzzi}, M. and {Pensotti}, S. and {Perrina}, C. and {Phan}, H.~D. and {Picot-Clemente}, N. and {Pilo}, F. and {Plyaskin}, V. and {Pohl}, M. and {Poireau}, V. and {Quadrani}, L. and {Qi}, X.~M. and {Qin}, X. and {Qu}, Z.~Y. and {R{\"a}ih{\"a}}, T. and {Rancoita}, P.~G. and {Rapin}, D. and {Ricol}, J.~S. and {Rosier-Lees}, S. and {Rozhkov}, A. and {Rozza}, D. and {Sagdeev}, R. and {Schael}, S. and {Schmidt}, S.~M. and {Schulz von Dratzig}, A. and {Schwering}, G. and {Seo}, E.~S. and {Shan}, B.~S. and {Shi}, J.~Y. and {Siedenburg}, T. and {Song}, J.~W. and {Tacconi}, M. and {Tang}, X.~W. and {Tang}, Z.~C. and {Tescaro}, D. and {Tian}, J. and {Ting}, Samuel C.~C. and {Ting}, S.~M.},
        title = "{Precision Measurement of Cosmic-Ray Nitrogen and its Primary and Secondary Components with the Alpha Magnetic Spectrometer on the International Space Station}",
      journal = {prl},
         year = 2018,
        month = aug,
       volume = {121},
       number = {5},
          eid = {051103},
        pages = {051103},
          doi = {10.1103/PhysRevLett.121.051103},
       adsurl = {https://ui.adsabs.harvard.edu/abs/2018PhRvL.121e1103A}
}

@ARTICLE{2023Univ....9...98S,
       author = {{Sinitsyna}, Vera G. and {Sinitsyna}, Vera Y.},
        title = "{Cosmic-Ray Acceleration in Supernova Remnants}",
      journal = {Universe},
         year = 2023,
        month = feb,
       volume = {9},
       number = {2},
          eid = {98},
        pages = {98},
          doi = {10.3390/universe9020098},
       adsurl = {https://ui.adsabs.harvard.edu/abs/2023Univ....9...98S}
}





5Author,

\end{document}